\documentclass[11pt,superscriptaddress,preprintnumbers, aps,prc,twocolumn, reprint,nofootinbib,longbibliography,tightenlines,floatfix]{revtex4-2}
\usepackage{amssymb,amsmath}
\usepackage{color}
\usepackage{hyperref}
\usepackage{listings}
\usepackage{physics}
\usepackage{amsmath}
\usepackage{graphicx}
\usepackage{bbold}
\usepackage{qcircuit}
\usepackage{bm}
\usepackage{amsfonts}
\usepackage{comment}
\usepackage{amsmath}
\usepackage{natbib}
\usepackage{multirow}
\usepackage{amsthm}
\usepackage{orcidlink}
\usepackage{forest}
\usepackage[normalem]{ulem}
\usepackage{subfig}
\usepackage{soul}

\newcommand{\R}{\mathbb{R}}
\newcommand{\C}{\mathbb{C}}
\newcommand{\refe}[1]{\textbf{$[$#1$]$}}
\newcommand{\meanvalue}[1]{\langle #1 \rangle}
\newcommand{\abv}[1]{\left| #1 \right|}
\renewcommand{\appendixname}{APPENDIX}
\usepackage[normalem]{ulem}

\hypersetup{
	unicode=false,          
	pdftoolbar=true,        
	pdfmenubar=true,        
	pdffitwindow=false,     
	pdfstartview={FitH},    
	pdfauthor={},     
	colorlinks=true,       
	linkcolor=blue,          
	citecolor=blue,        
}

\graphicspath{{figures/}}
\newcommand{\ut}{\ensuremath{\frac{\partial u}{\partial t}}}
\newcommand{\uxx}{\ensuremath{\frac{\partial^2 u}{\partial x^2}}}
\newcommand{\Qt}{\ensuremath{\frac{dQ}{dt}} }
\newcommand{\fixme}{{\bf **FIXME**}}
\renewcommand{\labelitemi}{$\vcenter{\hbox{\tiny$\bullet$}}$}

\definecolor{dkgreen}{rgb}{0,0.6,0}
\definecolor{gray}{rgb}{0.5,0.5,0.5}
\definecolor{mauve}{rgb}{0.58,0,0.82}

\newcommand{\blu}{\color{blue}}
\begin{document}
	\begin{figure}
		\vskip -1.cm
		\leftline{\includegraphics[width=0.25\textwidth]{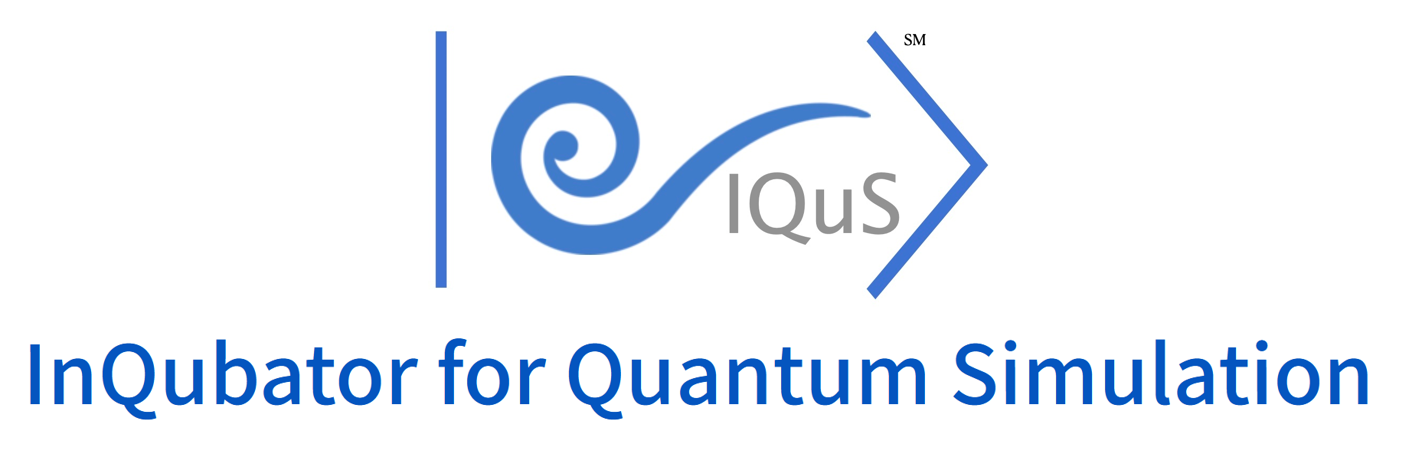}}
		\vskip -2.cm
	\end{figure}
	
	\preprint{Preprint number: LLNL-JRNL-865762-DRAFT; IQuS@UW-21-080}
	\hspace{0.1cm}

	\title{Evaluation of phase shifts for non-relativistic elastic scattering using quantum computers}
	
	\newcommand{\iqusfil}{InQubator for Quantum Simulation (IQuS), Department of Physics, University of Washington, Seattle, Washington 98195, USA}
	\newcommand{\llnlafil}{Lawrence Livermore National Laboratory, P.O. Box 808, L-414, Livermore, California 94551, USA}
	
	\newcommand{\utrentoafil}{Physics Department, University of Trento, Via Sommarive 14, I-38123 Trento, Italy}
	\newcommand{\infnafil}{INFN-TIFPA Trento Institute of Fundamental Physics and Applications, Via Sommarive 14, I-38123 Trento, Italy}
	
	\author{Francesco~Turro \orcidlink{0000-0002-1107-2873}}
	\thanks{fturro@uw.edu} 
	\affiliation{\iqusfil}
	
	
	\author{Kyle ~A.~Wendt\orcidlink{0000-0002-3428-6479}}
	\affiliation{\llnlafil}
	
	\author{Sofia Quaglioni\orcidlink{0000-0002-7512-605X}}
	\affiliation{\llnlafil}
	
	\author{Francesco Pederiva\orcidlink{0000-0002-7242-0042}}
	\affiliation{\utrentoafil}\affiliation{\infnafil}
	
	\author{Alessandro Roggero\orcidlink{0000-0002-8334-1120}}
	\affiliation{\utrentoafil}\affiliation{\infnafil}

	\begin{abstract}
		
		Simulations of scattering processes are essential in understanding the physics of our universe. 
		Computing relevant scattering quantities from ab initio methods is extremely difficult on classical devices because of the substantial computational resources needed.  
		This work reports the development of an algorithm that makes it possible to obtain phase shifts for generic non-relativistic elastic scattering processes on a quantum computer. This algorithm is based on extracting phase shifts from the direct implementation of the real-time evolution.
		The algorithm is improved by a variational procedure, making it more accurate and resistant to the quantum noise.
		The reliability of the algorithm is first demonstrated by means of classical numerical simulations for different potentials, and later tested on existing quantum hardware, specifically on IBM quantum processors.

	\end{abstract}

	\maketitle
	
	\section{Introduction}
	
	Many-body scattering processes play a key role in a large variety of physics phenomena ranging from nuclear reactions and decays that drive stellar evolution~\cite{astrophysics_1,astrophyscis_review2,RevModPhys_83_195} to electron-phonon scattering underlying superconductivity in materials~\cite{RevModPhys_89_015003,electron_scattering2}. Performing accurate simulations of scattering experiments for a generic quantum many-body system remains a challenging problem in theoretical physics. These simulations are extremely computationally demanding when the interactions between the constituents are non-perturbative, as in the case of nuclear physics. Typically, the classical computational resources required to perform such simulations grow exponentially with the number of particles. Moreover, many scattering processes involve fermions, which in classical methods often leads to the notorious fermion sign problem~\cite{troyer2005}, increasing the computational time needed to perform the simulations. Classical first principles (i.e., ab initio) simulations of nuclear reactions are currently limited to processes involving light elements such as those occurring during the Big Bang or in the proton-proton (pp) reaction chain in our Sun~\cite{Hupin:2019,Hebborn:2022,Kravvaris:2023} and, even in the exascale computing era, may remain limited to the description of $\alpha$ capture reactions~\cite{osti_1906717}.
	
	Since Feynman's original proposal in the 1980~\cite{feynman}, quantum computers have emerged as an interesting alternative to simulate quantum systems efficiently, in part thanks to the possibility to encode an exponential Hilbert space using a linear number of qubits. In particular, the simulation of real-time dynamics, and thus scattering processes, of locally interacting many-body systems was recognized early on as a promising application of digital quantum technologies~\cite{Lloyd_1996}. Recent years have witnessed a dramatic growth in the capabilities of quantum devices and multiple calculations have now been demonstrated on more than 100 qubits~\cite{100_qubits_0,100_qubits_1,100_qubits_2,100_qubits_3,100_qubits_4,100_qubits_5}, using different architectures ranging from trapped ions~\cite{Bruzewicz2019} to
	superconducting devices~\cite{huang2020superconducting, Bianchetti2010,wang2022high,wallraff2004strong,chiorescu2004coherent}
	and neutral atom arrays~\cite{Henriet2020quantumcomputing,browaeys2020many,Barredo2020,bluvstein2022quantum}. It is therefore timely and desirable to develop digital quantum algorithms for simulating scattering processes. 
	
	A variety of quantum algorithms have been proposed in the past for the simulation of generic scattering processes, among which is the pioneering work by Jordan-Lee-Preskill (JLP)~\cite{JordanLeePreskill,jordan2019quantum} for simulation of inelastic processes in scalar quantum field theory. In the context of nuclear physics, other quantum algorithms for dynamical simulations have been proposed, including the calculations of the inclusive and semi-exclusive scattering cross sections of nuclei to electroweak probes functions~\cite{roggero2019dynamic,roggero2020quantum,roggero_git,Du2021,baroni2022nuclear,Hartse_2023,kiss2024quantum}, radiative processes~\cite{bedaque2022radiative}, calculations of 
	Green’s functions~\cite{Endo2020,Anthony_scattering,Tong2021}, applications of the rodeo algorithm~\cite{wang2024nuclear}, hybrid quantum-classical methods to simulate semi-classical dynamics~\cite{turro2023quantumclassical} and inelastic scattering~\cite{Du2021,inelastic_scattering_Mueller} among others (see~\cite{Klco_2022,Miessen_2022,Bauer_2023,Bauer_2023b} for recent reviews).
	
	This work presents a quantum algorithm for a direct evaluation of phase shifts for a generic elastic scattering process, allowing the calculations of angular and total cross section. A variety of classical methods are available to simulate many-body scattering, from the famous  L{\"u}sher's formula~\cite{Lusher1,Lusher2,Lusher3,Lusher4}, R-matrix theory~\cite{descouvemont2010r,RevModPhys.30.257}, Monte-Carlo based methods~\cite{PhysRevLett.99.022502,PhysRevLett.116.062501,elhatisari2015ab}, optical potentials~\cite{PhysRevC.93.034619,PhysRevC.97.034619,PhysRevC.98.044625,PhysRevLett.123.092501} to No-Core Shell Models  with Continuum ~\cite{PhysRevC.87.034326,PhysRevLett.110.022505}. For two body scattering one can also apply direct integration schemes such as the Numerov method~\cite{numerov1927note}. Quantum algorithms for this problem have already been proposed recently~\cite{Gustafson2021,sharma2023scattering}. Our method provides an alternative scheme useful in computing phase shifts for elastic scattering in both single and multi-channel cases.
	
	We describe the main steps of the proposed algorithm in Sec.~\ref{sec:realtime} together with classical benchmark calculations showing its validity. In Sec.~\ref{sec:variational} we describe a variational variant of the approach which could be useful for implementation on near term noisy devices. Moreover, we also make the quantum algorithm noise-resistant to the decoherence of the processors with the variational method. 
	We present results of simulations on classical computers of the proposed algorithm in Sec.~\ref{sec:classical_results} using a variety of Hamiltonians and basis sets. In Sec.~\ref{sec:results_quantum_hardware} we report the results obtained from a direct implementation of our scheme on superconducting quantum processors from IBM~\cite{ibm,Qiskit} showing the impact of error mitigation techniques. Finally, we present our summary and conclusions in Sec.~\ref{sec:conclusions}.
	
	\section{The general real-time algorithm \label{sec:realtime}}

	Generically speaking, the theoretical description of a time-dependent scattering process formally consists of three essential steps. First, the system is prepared in a given initial state.  Next, the system is evolved in time guided by the many-body Hamiltonian. Finally the system is measured, potentially after contraction with a target state, to extract the physical observables.  In this work, we follow this archetypical approach to scattering, performing each of these steps as operations on a quantum device.
	
\begin{figure}[t]
		
		\includegraphics[width=\columnwidth]{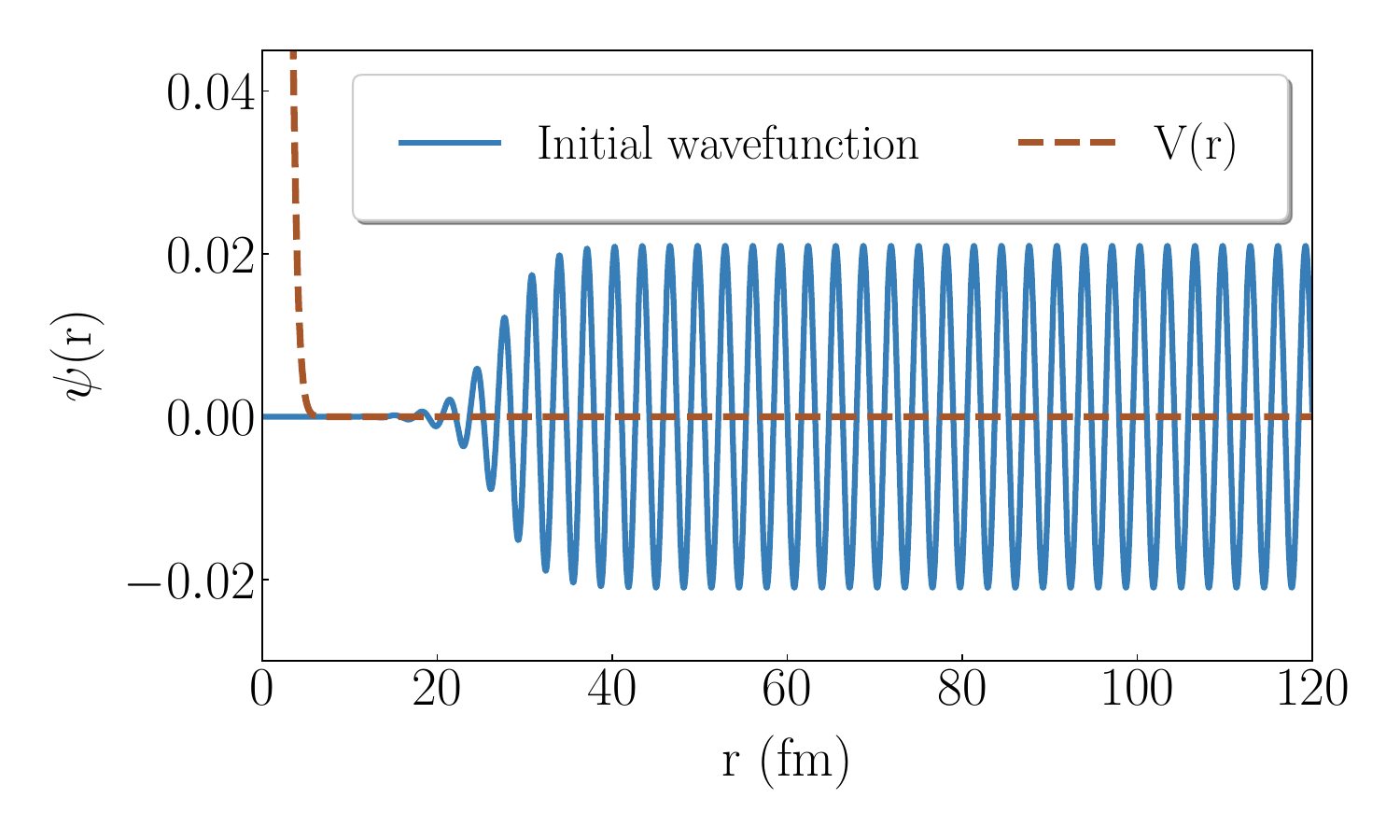}
		\caption{Example of an initial state of the real-time approach. The blue line represents the initial wave function, given by $\psi(r)= \frac{1}{1+e^{-\frac{(r-r_0)}{\Gamma}}} \sin(k_{\rm in} r)$ with $k_{\rm in}=2.12$ fm$^{-1}$, $\Gamma=2.5$ fm, and $r_0=28$ fm.  
			The dashed line represents the Gaussian potential $V(r)=V_0e^{-\frac{r^2}{\sigma^2}}$ with $V_0=1$ MeV and $\sigma=2$ fm.}
		\label{fig:initial_wave}
	\end{figure}

	For this first exploratory work,  we make two simplifications.  First, we work with a system of relative coordinates $\vec{r}$ between the particles. Second, we only consider central potentials $V(r)$ that depend on the magnitude $r$ of the relative coordinates $\vec{r}$.  Under these conditions, the wave function can be cast as an expansion of partial waves defined by their angular momentum $L$  
	\begin{equation}
		\Psi(\vec{r})=\sum_{L=0}^{\infty}\sum_{m=-L}^L A_{L,m}\frac{u_L(r)}{r}Y_L^m(\theta,\phi),
	\end{equation}
	where the $Y_L^m$ are the standard spherical harmonics and the radial functions $u_L$ are the solutions of the radial Schr\"{o}dinger equation,
	\begin{equation}
		\left\{\frac{\hbar^2}{2\mu}\frac{d^2}{dr^2}+\left[E-V(r)-\frac{\hbar^2}{2\mu}\frac{L(L+1)}{r^2}\right]\right\}u_L(r)=0,
	\end{equation}
	where $\mu$ is the reduced mass of the system.
	Since our goal is to extract the scattering phase shifts, it is advantageous to initialize the system in a state as close as possible to an incident plane wave. For instance, taking a plane wave of momentum 
	\begin{equation}
		k_{\rm in}=\sqrt{\frac{2\mu E}{\hbar
				^2} }  \label{eq:kini} 
	\end{equation}
	traveling in the $z$ direction, we have
	\begin{equation}
		\begin{split}
			\Psi_{in}(\vec{r})&=e^{ik_{\rm in}r\cos(\theta)}\\
			&=\sum_{L=0}^{\infty}a_L\,i^L(2L+1)j_L(k_{\rm in}r)P_L(\cos(\theta))\;, \label{eq:partialwave_exp}
		\end{split}
	\end{equation}
	where $j_L$ are the spherical Bessel functions of the first kind, $P_L$ the Legendre polynomials, and $\theta$ indicates the polar angle. We work in a finite volume, and therefore we find it convenient to modify an initial wave packet by multiplying the plane wave by a suitable filter function designed to suppress the amplitude in the interaction area, where the potential $V(r) \neq 0$. For example, a possible choice of the filter function can be the following,
	\begin{equation}
		f(r)=\frac{1}{1+e^{-\frac{(r-r_0)}{\Gamma}}} \,,\label{eq:filter_fermi}
	\end{equation}
	where $r_0$ corresponds to the boundary of the interaction region, and the parameter $\Gamma$ controls the width of the filter. An example of such a state for a central potential, with momentum $k_{\rm in}=2.12$ fm$^{-1}$ and angular momentum $L=0$, is shown as the blue line in Fig.~\ref{fig:initial_wave}. The brown dashed line instead corresponds to a central Gaussian potential.

	After preparing the truncated plane wave, the real-time evolution is implemented by applying the propagator $e^{-it\hat{H}}$, where $\hat{H}$ is the radial Hamiltonian:
	\begin{equation}
		\label{eq:radial_ham}
		\hat{H}= -\frac{\hbar^2}{2\mu}\frac{d^2}{dr^2} + V(r)+\frac{\hbar^2}{2\mu}\frac{L(L+1)}{r^2}\;.
	\end{equation}
	
	When the propagation is implemented for a sufficiently long time $t$, we expect the  evolved wave function to converge to the solution of the  Lippmann–Schwinger equation~\cite{griffiths,weinberg_book} that, for $r$ larger than the range of the potential, can be written as:
	\begin{equation}
		\lim_{t\to\infty} u_L(r,t)\propto kr[\cos(\delta_L) j_L(k_{\rm in}r)-\sin(\delta_L) n_L(k_{\rm in} r)]\;,
	\end{equation}
	where $n_L$ is spherical Bessel functions of the second kind, and $\delta_L$ is the phase-shift. In the $r\rightarrow\infty$ limit, the Bessel functions can be further approximated by sines and cosines, giving the asymptotic expression for the radial solution,
	\begin{equation}
		\lim_{t\to\infty}\lim_{r\to\infty} u_L(r,t)= A_L\,\sin(k_{\rm in}\,r+\frac{\pi\,L}{2}+\delta_L) 
		\,.
		\label{eq:scatteringeq}
	\end{equation}
	
	Furthermore, the energy conservation given by the unitarity of the propagator also guarantees that the momentum of the asymptotic solution is still equal to $k_{\rm in}$ at the time of detection.
	
	In order to extract information about the phase-shifts $\delta_L$ on a quantum device, we can measure the overlap between the evolved state $\psi_L(t)$ and a detector state, defined as the initial plane wave in an interval $r\in[r_1,r_2]$ in the asymptotic region, where we set 
	\begin{equation}
		\label{eq:r2_def}
		r_2=r_1+\frac{2\pi}{k_{\rm in}} n_d\,
	\end{equation}
	with $n_d$ an integer number. Hence, we define the detector state as
	\begin{equation}
		\ket{\phi_D}= c_D \,{\rm Box}(r,r_1,r_2) \,j_{L}(k_{\rm in} r)\,, \label{eq:detector_state_box}
	\end{equation}
	where $c_D$ indicates a normalization constant, and $\rm{Box}(r_1,r_2)$ the Box function between $r_1$ and $r_2$
	\begin{equation}
		{\rm Box}(r,r_1,r_2) = \theta(r-r_1)-\theta(r-r_2)\;,
	\end{equation}
	with $\theta(x)$ the Heaviside  $\theta$ function, and we consider asymptotically large values for $r_1$.
	
	Therefore, we can write the overlap between $\phi_D$ and $\psi_L(t)$ as,
	\begin{equation}
		\begin{split}
			O(t) & =\bra{\phi_D} \ket{\psi_L(t)}\\
			&\approx  \alpha_L \int_{r_1}^{r_2} dr\sin(k_{\rm in}\,r+\frac{\pi\,L}{2})\\
			&\quad\quad\quad\quad\quad\times \sin(k_{\rm in}\,r+\frac{\pi\,L}{2}+\delta_L)\\
			&=\frac{r_2-r_1}{2}\alpha_L\cos(\delta_L)\\
			&\qquad +\frac{\sin(k_{\rm in}(r_1-r_2))}{2k_{\rm in}}\alpha_L\cos\left(k_{\rm in}(r_1+r_2)+\delta_L\right) \\
			&=\frac{r_2-r_1}{2}\alpha_L\cos(\delta_L)\,,
			\label{eq:Prob_delta_0}
		\end{split}
	\end{equation}
	where in the second line we used Eq.~\eqref{eq:scatteringeq} and introduced $\alpha_L=c_DA_L$. In the last line, we use the definition of $r_2$ from Eq.~\eqref{eq:r2_def}. As a consequence, the phase shift $\delta_L$ can be obtained from the overlap probability $P=\left|O(t)\right|^2$.
	
	We can initialize the system in a state corresponding to a single component of $\ket{\Psi_{\rm in}}$ with a fixed angular momentum $L$ by representing only the radial part on a spatial grid with $2^M$ points,
	\begin{equation}
		\ket{\Phi_{L}(0)}=\mathcal{N}\sum_{m=0}^{2^M}(k_{\rm in}r_m)j_L(k_{\rm in} r_m)\ket{m}\;,
	\end{equation}
	where $\mathcal{N}$ is the appropriate normalization factor.
	The value of initial plane wave momentum $k_{\rm in}$ (see Eq.~\eqref{eq:kini}) is chosen to be an eigenvalue of the kinetic energy operator in order to reduce numerical noise during the real-time evolution (the initial wave function is close to the kinetic eigenstate). 
	
	We then perform the time evolution with only the radial Hamiltonian $\hat{H}_{\rm eff}$ from Eq.~\eqref{eq:radial_ham} and we compute the overlap probability with the detector state $\ket{\phi_D}$,

	\begin{equation}
		P=\lim_{t\to\infty}\left|\langle\Phi_D\vert\psi_L(t)\rangle \right|^2=c^2_L\cos^2(\delta_L)\;, 
		\label{eq:overlap_prob_delta_L}
	\end{equation}
	with $c_L$ being a proportionality constant that depends on the choice of $r_1$ and $r_2$ in the detector. We stress that we can obtain only the absolute value of $\delta_L$ because $\pm \delta_L$ has the same overlap probability value. Sec.~\ref{sec:variational} presents an improvement of the quantum algorithm that also allows learning the correct sign of the phase shift.
	
	The coefficient $c_L$ indicates the change in normalization from the detector volume, and it can be approximated by the overlap probability between the initial state and the detector state (considering it far away). Its value is obtained from
	\begin{equation}
		c_L^2=\int_{r_1}^{r_2} dr\,r^2\, \frac{1}{c_{\rm init}} j_L(k_{\rm in}r) \frac{1}{c_{\rm dect}}j_L(k_{\rm in}r) = {\frac{{c_{\rm dect}}}{c_{\rm init}}}\,,
	\end{equation}
	where $c_{\rm init}$ and $c_{\rm dect}$ parameters indicate the normalization of the initial and detector states, respectively.  They are given by 
	\begin{equation}
		\begin{split}
			c_{\rm dect}&=\sqrt{\int_{r_1}^{r_2}dr \,r^2 j_L^2(k_{\rm in}r)} \\
			c_{\rm init}&=\sqrt{\int_{0}^{R_{max}}dr \,r^2 f^2(r) j_L^2(k_{\rm in}r)}\;,
		\end{split}
	\end{equation}
	where $R_{max}$ is the size of the simulation box.

	In this work, we choose to use a uniform spatial grid where one can naturally find locations $r_1$ and $r_2$ in the asymptotic limit trivially.  Using the formal tools of a discrete variable representation, this entire analysis can be generalized to any basis representation where the basis follows a typical three term recurrence relationship~\cite{Bulgac:2013,Binder:2016}. In this work, we calculate phase shifts for a central potential and for the angular momentum $L=0$. However, the same quantum algorithm can be applied for couple-channel scattering, where the Hamiltonian mixes different angular momenta.

	The SWAP test~\cite{swap_test} can easily evaluate the overlap probability of Eq.~\eqref{eq:overlap_prob_delta_L} on a quantum processor. It requires $2\,n_s+1$ qubits, where $n_s$ qubits are needed to map the evolved state, $n_s$ more to describe the free plane wave, and an extra qubit serves as an ancillary one. Nevertheless, this implementation might be hard in today's processors with few available qubits. Computationally cheaper alternatives are possible, using more sophisticated implementations of the SWAP test~\cite{Cincio_2018} or randomized measurements~\cite{Elben_2022}. We adopt an alternative approach where we implement an operator $\hat{D}^\dagger$, where $\hat{D}$ is defined as the operator that reinitializes the detector state of Eq.~\eqref{eq:detector_state_box},
	\begin{equation}
		\ket{\phi_D} = \hat{D} \ket{0...0} \, ,
	\end{equation}
	where $\ket{0...0}$ indicates the initial state of the qubits.
	
	After applying the $\hat{D^\dagger}$ operator on $\ket{\psi(t)}$ and measuring all the qubits in the $\ket{0}$ state, we recover Eq.~\eqref{eq:overlap_prob_delta_L}. Indeed, using the definition of the $D$ operator, we can write
	\begin{equation}
		\begin{split}
			P &= \left|  \bra{0...0} D^\dagger \ket{\psi(t)}\right|^2 = \left| \left(D \ket{0...0}\right)^\dagger \ket{\psi(t)}\right|^2   \\
			&= \left| \bra{\phi_D}\ket{\psi(t)}\right|^2 = c_L^2\,\cos(\delta_L)^2\,.\\
		\end{split}
	\end{equation}

	\begin{figure}[t]
		\includegraphics[width=\columnwidth]{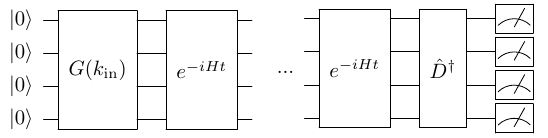}
		\caption{Scheme of the quantum circuits for computing phase shift. The gates $G(k_{\rm in})$ and $\hat{D^\dagger}$ are the gates that prepare and unprepare the initial and detector state, respectively. See the main text for details.}
		\label{fig:qc_truncation}
	\end{figure}
	\begin{figure}[t]	
		\includegraphics[width=\columnwidth]{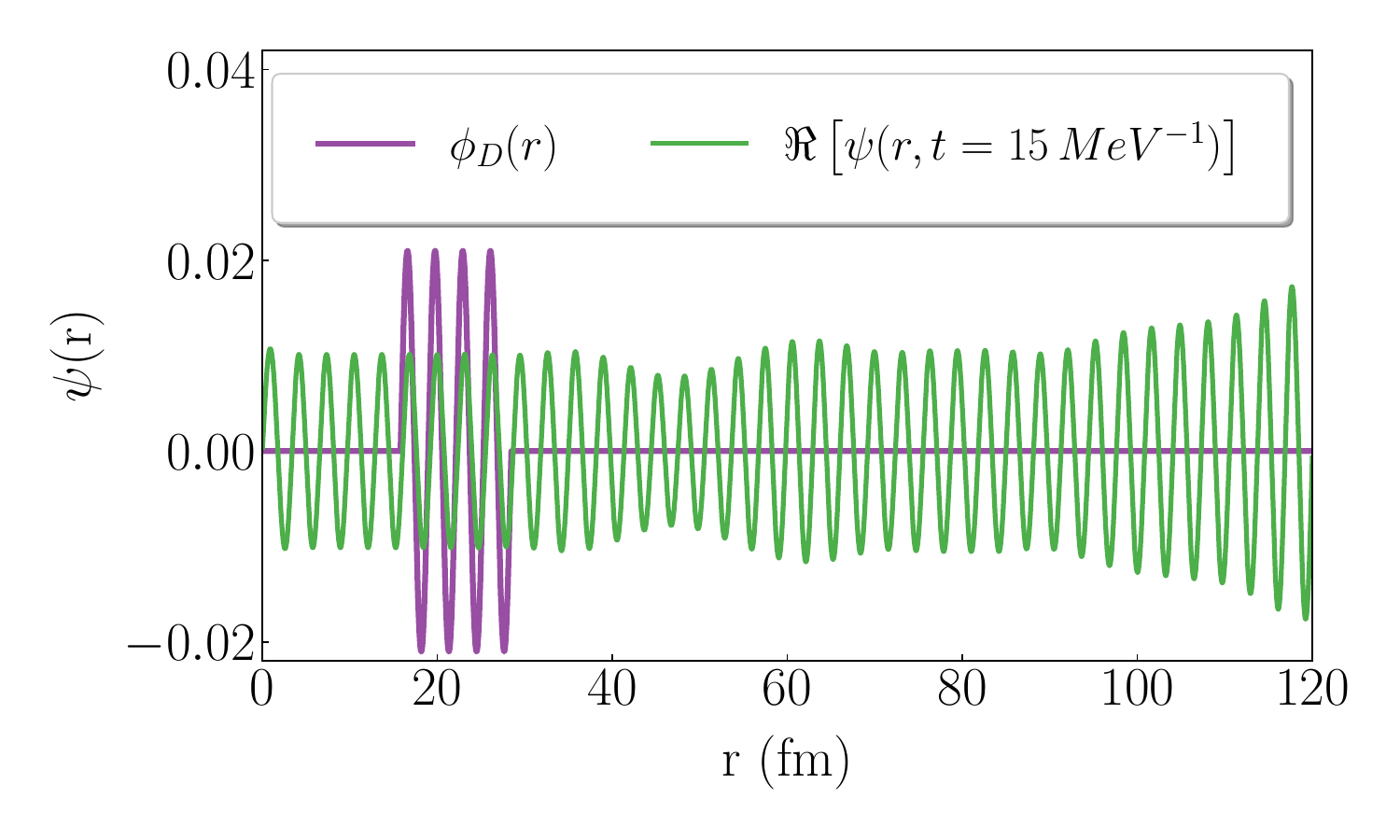}
		\caption{Plot of the real part of the evolved wave function (green line) at $t=15$ MeV$^{-1}$ for a scattering with a Gaussian potential. The purple line indicates the detector state.}
		\label{fig:wavefunction_delta_0_time_t}
	\end{figure}
	
\begin{figure}
		\centering
		\includegraphics[scale=0.8]{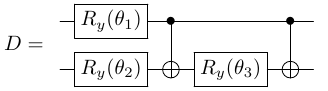}\\
		\includegraphics[scale=0.8]{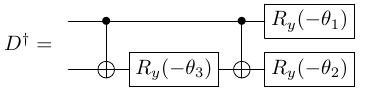}
		\caption{Example of how we can obtain the quantum circuit for the $D^\dagger$ (lower circuit) from one for $D$ (top circuit).}
		\label{fig:detector_qc_scheme}
	\end{figure}

	At the end, we can summarize the general scheme of the quantum circuit in Fig.~\ref{fig:qc_truncation}. We implement a gate, called $G(k_{\rm in})$, that prepares the quantum processor in the truncated wave function. Then, we evolve the state according to the system Hamiltonian, generally using the Trotter decomposition. At the end, we implement the detector gate, $D^\dagger$ and the probability that all qubits are in $\ket{0...0}$ state is equal to Eq.~\eqref{eq:overlap_prob_delta_L}. We highlight the fact that we can apply the same state-preparation algorithm for preparing $G(k_{\rm in})$ and $D^\dagger$. Indeed, we compute the quantum circuit for $D$ operator (this prepares the system in the detector state of Eq.~\eqref{eq:detector_state_box}). Then, the quantum circuit for $D^\dagger$ is obtained from the gates of $D$, flipping the sign of the angle of single-qubit rotations and implementing the gate starting from right side to left side.  Fig.~\ref{fig:detector_qc_scheme} shows an simple example with two qubits gate. We call the presented quantum algorithm method Time Evolution Phase Shift (TEPS).
	
	Different quantum state-preparation methods can be implemented for implementing $G(k_{\rm in})$ and $D^\dagger$ operators. A general quantum algorithm, that is also applied in this work, is given by Sec.~3 of Ref.~\cite{PhysRevA.102.012612} that efficiently prepares a real wave function. This algorithm is based on implementing a sequence of controlled-$R_y$ gates decomposed using Gray code, where the combination of $R_y$ angles gives the desired coefficients. This algorithm is enough to prepare a general plane wave because the partial wave expansion of the truncated initial plane wave and detector is real for even $L$ and purely imaginary for odd $L$. The extra $i$ factor for odd $L$ can be obtained implementing extra $CZ$ gates that adds the phase $i$ on qubits where one maps the odd $L$ waves.

	\subsection{Classical Tests}
	
	The validity of the TEPS algorithm was tested by expanding the Hamiltonian with the Gaussian and H-Ar Lennard-Jones potentials in a spatial lattice for $L=0$. The details of the adopted Hamiltonians can be found in Appendix~\ref{app:Hamiltonian}.
	
	We test the TEPS algorithm with a Gaussian potential ($V_0=1$, $\sigma=2$  and $\frac{\hbar^2}{2\mu}=1$) with an initial momentum $k_{\rm in}=1.99$ fm$^{-1}$. We use a finite box with $R_p=6000$ lattice points and a lattice spacing $a=0.02$ fm. 
	Fig.~\ref{fig:wavefunction_delta_0_time_t} shows the obtained snapshot for the evolution at time 15 MeV$^{-1}$ with the green line. The purple line represent the used detector state to compute the overlap probability.
	
\begin{figure}[b]
		
		\includegraphics[width=\columnwidth]{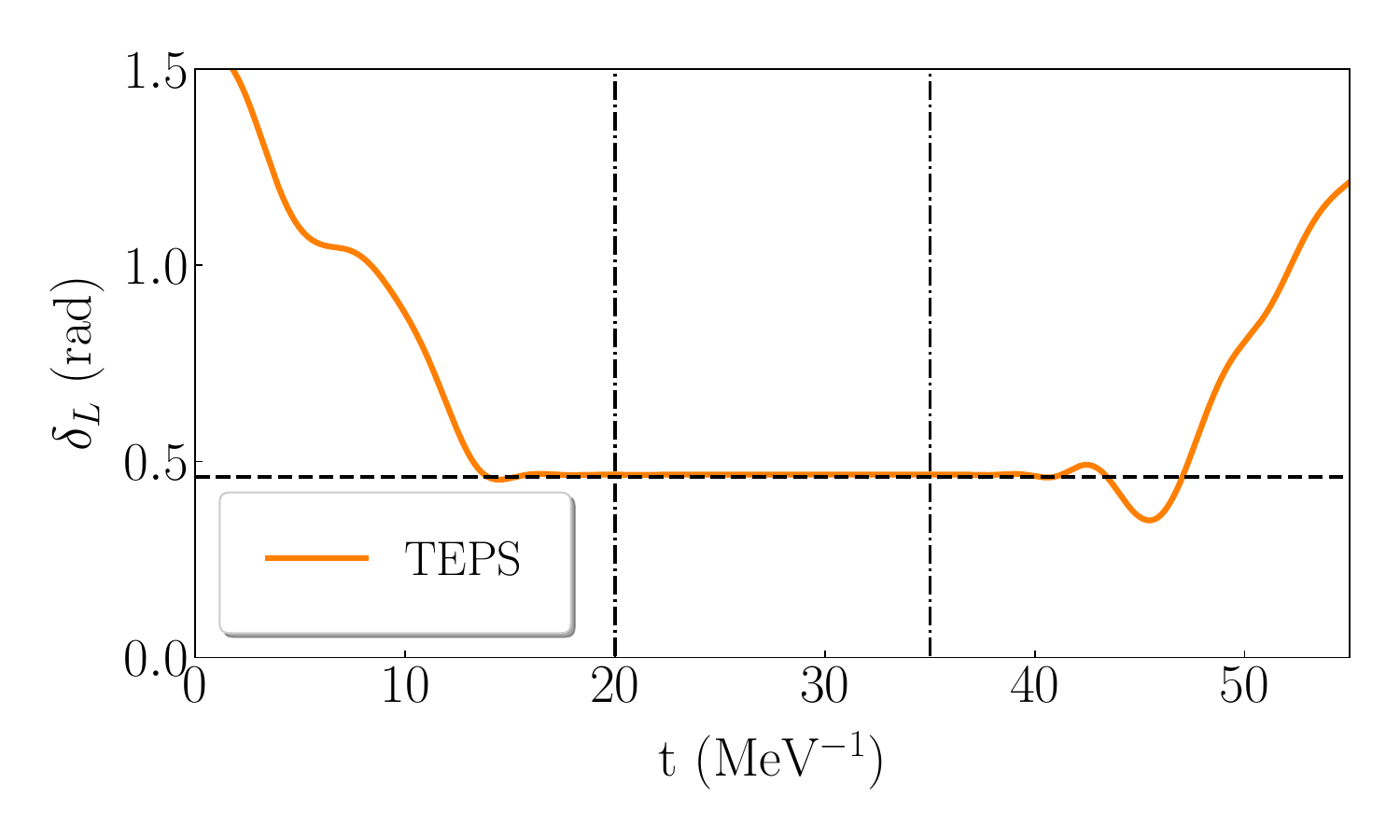}
		\caption{Estimation of the phase shift as a function of time $t$ for the scattering with a Gaussian potential. The dashed horizontal line corresponds to the exact phase shift. The vertical dotted-dashed lines delimit a possible interval for averaging the phase shift results. Numerical results are shown in Table~\ref{tab:phaseshift_probability}.}
		\label{fig:phaseshift_delta_0_time_t}
	\end{figure}

	Figure~\ref{fig:phaseshift_delta_0_time_t} shows the results for the phase shift results as a function of time. The dashed horizontal line indicates the exact phase shift calculated from the Numerov algorithm \cite{numerov1927note}. We can observe that between $t=24$ MeV$^{-1}$ and $t=34$ MeV$^{-1}$  (shown with vertical dotted lines), the phase shift values are constant. Averaging over this interval, we obtain $\delta_L= 0.46(3)$. 
	Such result is compatible within two standard deviations with the exact one, $\delta_L^{ex}=-0.47$. 
	
	We can also observe that after $t=40$ MeV$^{-1}$, the phase shift starts changing. This is due to border effects. Part of an evolved wave bounces on the infinite wall at the right end of the box, $R_{max}=120$ fm, and re-enters in the detector area.  The plateau interval can be extended by increasing the lattice points. Moreover, the initial time of the plateau (named $\tau$) can be empirically estimated by the passing time for a free wave with momentum $k_{\rm in}$  starting from $r_0$ position goes to $0$ and bounces back to the middle of the detector position $r_d$ ($r_d=\frac{r_1}{2}+\frac{r_2}{2}$).  This initial plateau time is then given by,
	\begin{equation}
		\tau\,\simeq\,\frac{1}{k_{\rm in}}\,\left(r_0+r_d\right) \,.  
	\end{equation}
	
	Table~\ref{tab:phaseshift_probability} illustrates the results obtained for different potentials and momenta. In all present lattice simulations, the number of lattice points is set to $R_p=6000$, the lattice spacing is $a=0.02$ fm for the Gaussian potential and  $a=0.02$ \r{A} for the Lennard-Jones potentials.
	
	\begin{table}[t]
		\caption{Phase shifts $\delta_L^{TEPS}$ computed by the TEPS algorithm, tabulated along the absolute value of exact phase shift $\delta_L^{ex}$ for the H-Ar scattering. The parameters of the Lennard-Jones potential used can be found in Ref.~\cite{Lennard_Jones_potential}}  
		\begin{tabular}{|c|cc|}
			\hline
			\multicolumn{3}{|c|}{Gaussian $V_0=1$ MeV $\sigma=2$ fm}\\
			\hline
			& $\left|
			\delta_L^{ex}\right|$& $\delta_L^{TEPS}$  \\
			\hline
			$k_{\rm in}=1.466$ fm$^{-1}$&0.646&
			0.65(3)\\
			$k_{\rm in}=1.73$ fm$^{-1}$& 0.540 & 0.54(3)\\ 
			$k_{\rm in}=2.252$ fm$^{-1}$& 0.4599
			&0.40(3)\\ 
			\hline
			\multicolumn{3}{|c|}{H-Ar Lennard-Jones}\\
			\hline
			& $\left|
			\delta_L^{ex}\right|$& $\delta_L^{TEPS}$  \\
			\hline
			$k_{\rm in}=0.602$ \r{A}$^{-1}$& 0.753&  0.76(3)\\
			$k_{\rm in}=0.8639$ \r{A}$^{-1}$&  0.543 & 0.54(3)\\
			$k_{\rm in}=0.995$ \r{A}$^{-1}$&  1.148 &
			1.14(3)\\
			
			\hline
		\end{tabular}
		\label{tab:phaseshift_probability}
	\end{table}
	
	Appendix~\ref{app:parameters_algorithm} discusses the effects of the truncation and lattice parameters in the phase shift results. From our analysis, the truncation effects can be mainly mitigated by increasing the lattice and increasing the $\Gamma$ value (filter function width) in the initial state.

	\section{Improvement with a Variational Approach}
	\label{sec:variational}
	
\begin{figure}
		\includegraphics[width=\columnwidth]{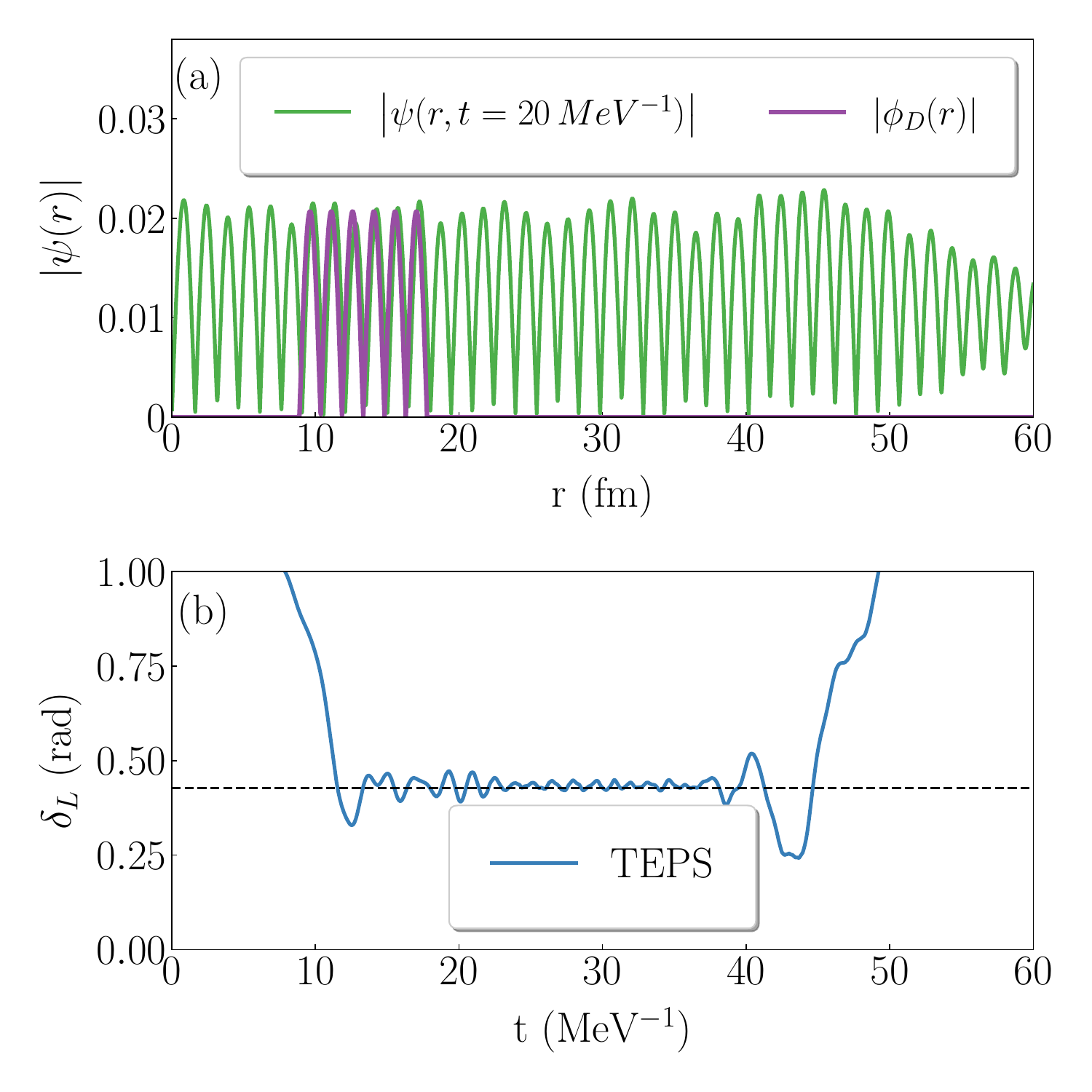}
		\caption{TEPS results for the phase shift of H-Ar scattering modeled by a Lennard Jones potential. Panel (a): wave function computed at $t=20$ MeV$^{-1}$ for $k_{\rm in}=2.12 \rm{ fm}^{-1}$. Panel (b): Value of the Phase Shift as a function of time.}
		\label{fig:messwave}
	\end{figure}

	In the previous section, we found a good compatibility between the TEPS phase shift results and the exact ones for general Hamiltonians. However, this agreement depends on the initial set-up of our simulation. Figure~\ref{fig:messwave} shows the phase shift computed for the scattering with a Gaussian potential ($V_0=1$ MeV, $\sigma=2$ fm.) starting with a filter width $\Gamma=0.0001$ fm\footnote{In Fig.~\ref{fig:phaseshift_delta_0_time_t} we use $\Gamma=2.5$ fm} . With this small value, we pick contributions from many undesired momenta, causing a large variance of the phase shift results around the plateaus region. If we take a snapshot of the evolution, as shown in Fig.~\ref{fig:messwave}a, we observe that the peak amplitudes are not stable, causing a big variance of the overlap probability.
	The green line shows the absolute value of the evolved wave function at time $t=20$ MeV$^{-1}$ and the purple line the used absolute value of the detector.  This causes a big  
	variance of the phase shift results shown in Fig.~\ref{fig:messwave}b. Moreover, from the analysis in Appendix~\ref{app:parameters_algorithm}, we observe the wrong lattice spacing, the lattice points or the initial width filter $\Gamma$ can contribute mostly on errors and variance of phase shift results. 
	
	\begin{table}[bt]
		\caption{Obtained results of the absolute value of Phase Shifts ($\delta_L$) computed by the difference between the peaks of the scattered wave function ($r_{full}$) and the free plane waves ($r_{free}$) for the scattering with a Gaussian potential for momentum $k_{\rm in}=2.12$ fm$^{-1}$. The exact phase shift is -0.428.}
		\begin{tabular}{|cc|c|}
			\hline
			$r_{free}$  [fm] &$r_{full} $  [fm] & $|\delta_L|$ [rad] \\
			\hline
			11.30 &	 9.64 & 0.378(13) \\
			12.80 &	11.12 &	0.420(13)\\
			14.30 &	12.60 & 0.463(13)\\
			15.76 &	14.08 & 0.420(13) \\
			17.24 & 15.56 & 0.420(13)\\
			18.74 & 17.04 & 0.463(13)\\
			20.22 & 18.52 & 0.463(13) \\
			21.68 &	20.00 & 0.420(13) \\
			23.16 & 21.48 & 0.420(13) \\
			24.64 &	22.96 & 0.420(13)\\
			26.14 & 24.46 & 0.420(13)\\
			27.64 & 25.94 & 0.463(13)\\
			29.12 & 27.42 & 0.463(13) \\
			30.56 & 28.90 & 0.378(13)\\
			32.04 & 30.38 & 0.378(13)\\
			33.54 & 31.86 & 0.420(13)\\
			35.02 &  33.34 & 0.420(13)\\
			36.50 &34.82 & 0.420(13)\\
			38.02 &36.30 & 0.505(13)\\
			39.52 & 37.78 & 0.547(13)\\
			\hline
		\end{tabular}
		
		\label{tab:phaseshift_wavepacket}
	\end{table}
	
	Nevertheless, in many cases, we can obtain the correct phase shift by looking at the difference between the peaks of the evolved state and the free plane wave. The phase differences of Fig.~\ref{fig:messwave} are reported in Table~\ref{tab:phaseshift_wavepacket}. We can notice that these phase results are close to the exact value inside three sigma.
	
	In order to reduce the TEPS sensitivity on noisy wave functions, we introduced a variational method that improves the algorithm. We found that such variational method makes the TEPS algorithm resistant to some noise sources of quantum processors, as discussed in Sec.~\ref{sec:results_quantum_hardware}.

	We modify Eq.~\eqref{eq:Prob_delta_0}, adding a fictitious phase $\delta_V$ to the free plane wave. In particular, the detector target wave function becomes $\sin(k_{\rm in}\,r+\frac{L\pi}{2}+\delta_V)$. The new overlap probability becomes 
	\begin{equation}
		\begin{split}
			P(\delta_V)  &= \int_{r_1}^{r_2} dr \sin(k_{\rm in}\,r+\frac{\pi\,L}{2}+\delta_L) \\ &\times \qquad\qquad\quad\quad\sin(k_{\rm in}\,r+\frac{\pi\,L}{2}+\delta_V) \\
			&= b_0\,\cos(\delta_V-\delta_L)^2        \,, \label{eq:P_delta_V}
		\end{split}
	\end{equation}   
	where $r_1=\frac{(2\pi n_i+\delta_V)}{k_{\rm in}}$ and $r_2=\frac{(2\pi n_f+\delta_V)}{k_{\rm in}}$ and $n_i,n_f \in \mathbb{Z}$.
	
	From Eq.~\eqref{eq:P_delta_V}, the correct phase shift $\delta_L$ is computed by the value $\delta_V=\delta_V^{max}$ that maximizes $P(\delta_V)$. Moreover, we also recover the correct sign of the phase shift.
	Using this procedure, $P(\delta_V)$ does not depend on the wave function amplitudes. Therefore, even in the case of the evolution of noisy wave functions (Fig.~\ref{fig:messwave}), we obtain the correct phase shift value. An analytic proof is presented in Appendix~\ref{app:variational_proof}. We named this variational approach the Variational Time Evolution Phase Shift (V-TEPS) algorithm.

	\section{Classical results for the Variational method \label{sec:classical_results}}
	
	In this section we show how the V-TEPS algorithm can be used in a classical calculation using different basis sets: a spatial lattice and a momentum basis set. Once again the results are given for scattering on a Gaussian and on a Lennard-Jones potentials.

	\subsection{Spatial Lattice calculations \label{subsec:lattice_variational}}

\begin{figure}[th]
		
		\includegraphics[width=\columnwidth]{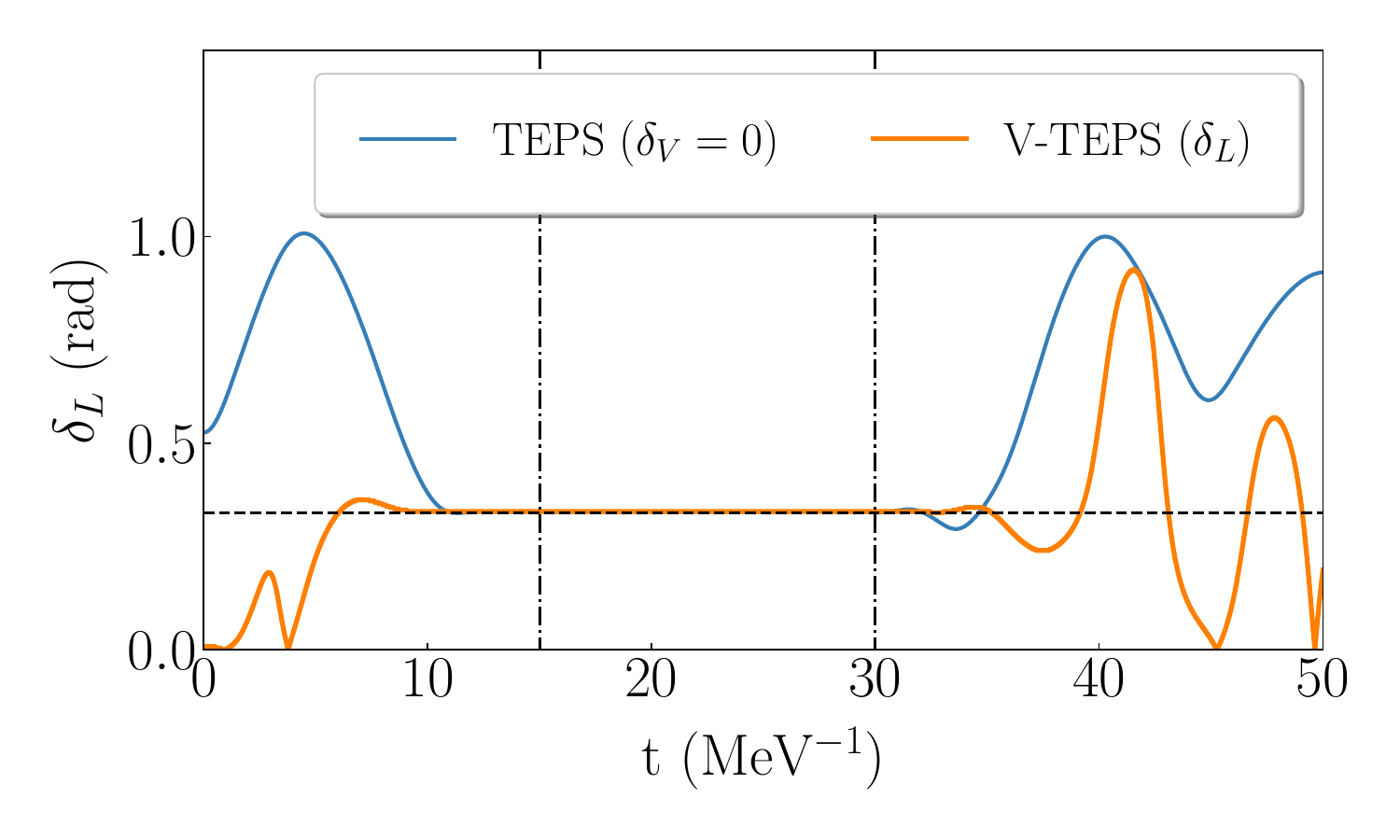}
		\caption{Phase shift results for the Gaussian potential at $k_{\rm in}= 2.67$ fm$^{-1}$. The blue line represents the results computed from the TEPS method (case $\delta_V=0$). The orange and blue line indicate the absolute value of the V-TEPS and TEPS results, respectively. The vertical dashed line represents the interval used to average the phase shift. The obtained result is reported in Table~\ref{tab:lattice_variational} }
		\label{fig:variational_Gaussian}
	\end{figure}

\begin{figure}[th]
		\includegraphics[width=\columnwidth]{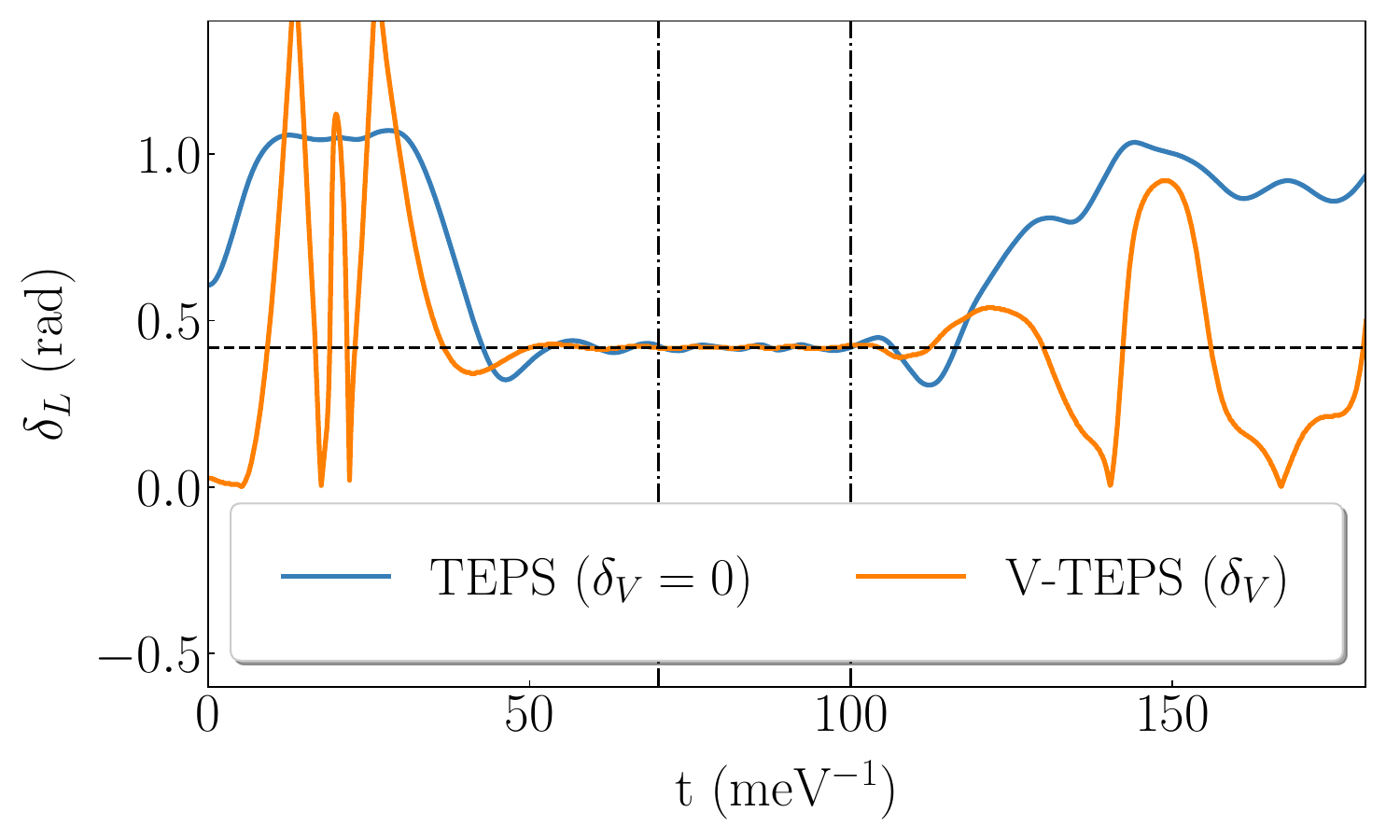}
		\includegraphics[width=\columnwidth]{Variational_lattice_leonard_jones.pdf}
		\caption{Phase shift results for the Lennard-Jones potential at $k_{\rm in}=0.66$ \r{A}$^{-1}$.The orange and blue line indicate the absolute value of the V-TEPS and TEPS results, respectively. The vertical dashed line represents the interval used to average the phase shift. The obtained result is reported in Tab.~\ref{tab:lattice_variational}.}
		\label{fig:variational_LJ}
	\end{figure}

	Figure~\ref{fig:variational_Gaussian} shows the phase shift computed for $k_{\rm in}=2.67$ fm$^{-1}$ and the Gaussian potential ($V_0=1$ MeV, $\sigma=2$ fm). Figure~\ref{fig:variational_LJ} shows instead the results for $k_{\rm in}=0.66$ \r{A}$^{-1}$ and the Lennard-Jones potential. In both plots, the blue lines represent the phase shift results as a function of time obtained with TEPS (see Sec.~\ref{sec:realtime}), while the orange line indicates the phase shift computed when we apply the variational method. The plotted results are obtained from the values of the parameter $\delta_V$ that maximizes the overlap probability. Figures show that the variational method gives a more stable estimate for the phase shift. Also, the interval over which the phase shift is constant becomes longer with the variational method.
	
	Table~\ref{tab:lattice_variational} reports the numerical values of the V-TEPS phase shift for both the Gaussian and Hydrogen-Argon (H-Ar) Lennard Jones potentials and for different momenta.

	\begin{table}[ht]
		\caption{Phase shift (PS) results for the Gaussian and Lennard-Jones potential for different momenta. The TEPS column results are obtained using the original algorithm of Sec.~\ref{sec:realtime}. The V-TEPS results are obtained by implementing the algorithm of Sec.~\ref{sec:variational}.}
		\begin{tabular}{|c|c|c|c|}
			\hline
			\multicolumn{4}{|c|}{Gaussian $V_0=1$ MeV $\sigma=2$ fm}\\
			\hline
			$k_{\rm in}$ & Exact $\delta_L$& TEPS & V-TEPS \\
			\hline
			
			$1.72$ fm$^{-1}$& -0.54& 0.54(3)& -0.54(2)\\
			$2.12$ fm$^{-1}$  &-0.43 & 0.43(3)& -0.43(2)\\
			$2.67$ fm$^{-1}$ &-0.33 & 0.33(3)& -0.33(2)\\
			\hline
			\multicolumn{4}{|c|}{Gaussian $V_0=2$ MeV $\sigma=4$ fm}\\
			\hline
			$k_{\rm in}$ & Exact $\delta_L$& TEPS & V-TEPS \\
			\hline
			$1.334$ fm$^{-1}$ &
			-0.16&  0.16(4) & -0.19(3)\\
			
			$1.86$ fm$^{-1}$ &
			0.98 &  0.97(3) & 0.98(2)\\
			$2.51$ fm$^{-1}$ &
			-1.49 &  1.50(3) & -1.50(2)\\
			\hline
			
			\multicolumn{4}{|c|}{H-Ar Lennard-Jones}\\
			\hline
			$k_{\rm in}$ & Exact $\delta_L$& TEPS & V-TEPS \\
			\hline
			$0.67$ \r{A}$^{-1}$&  0.42 & 0.42(3) & 0.42(2) \\ 
			$1.19$ \r{A}$^{-1}$ & 1.13 & 
			1.13(3) & 1.13	(2)\\
			$1.58$ \r{A}$^{-1}$ & -0.49 & 0.48(3) & -0.48(2) \\
			\hline
		\end{tabular}
		
		\label{tab:lattice_variational} 
	\end{table}
	
\begin{figure}[th]
		
		\includegraphics[width=\columnwidth]{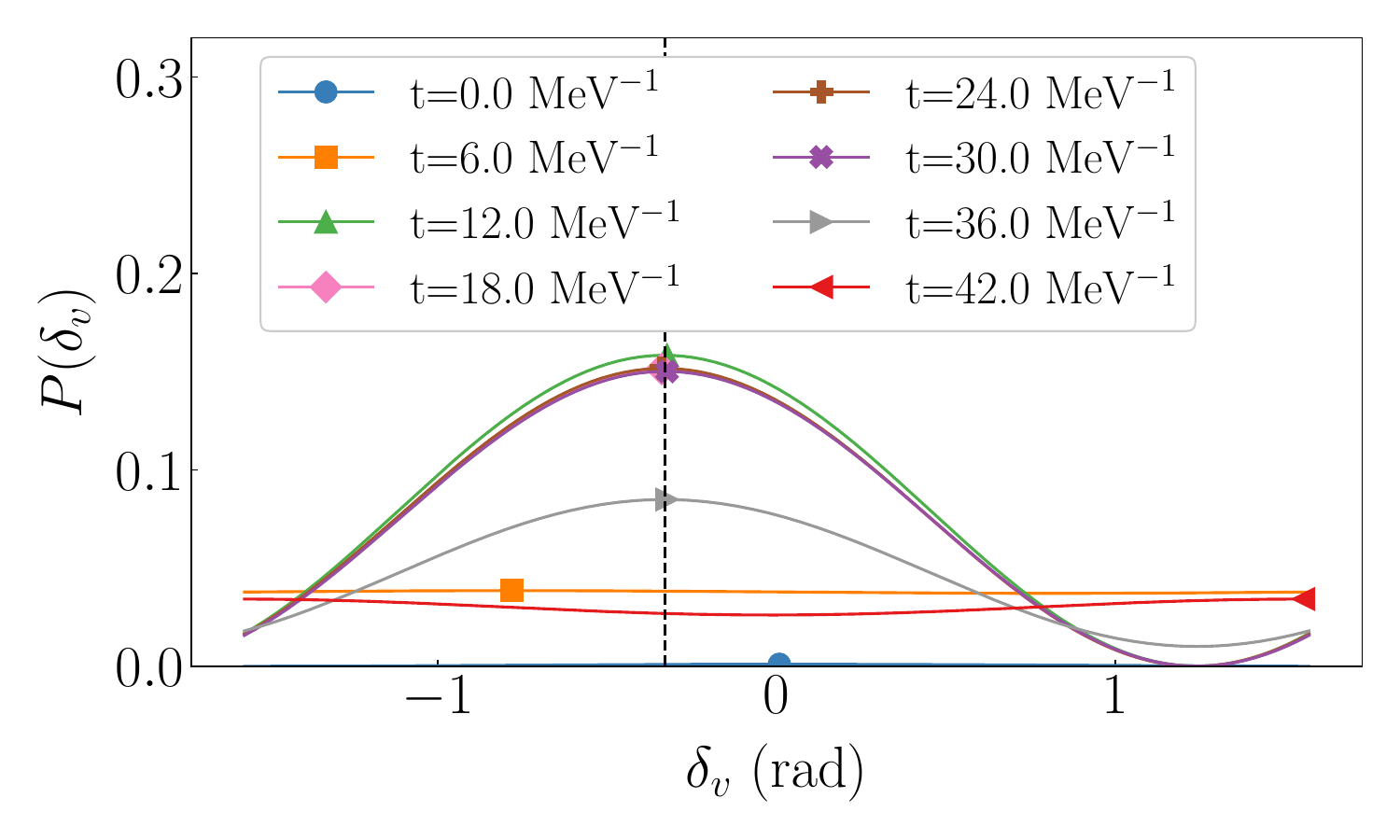}
		\caption{Probability as a function of the detector phase $\delta_V$ for different time steps. The dashed vertical line indicates the exact phase shift value. Each symbol indicates the position of the maximum value. The vertical dashed line shows the exact phase shift.}
		\label{fig:variational_prob_time}
	\end{figure}

	It is also interesting to look at the overlap probability dependence on $\delta_L$ for different times. The lines in Fig.~\ref{fig:variational_prob_time} illustrate the obtained overlap probability as a function of $\delta_V$ for the same system in Fig.~\ref{fig:variational_Gaussian} for different times. We can observe that from time $t=12$ MeV$^{-1}$ to $t=30$ MeV$^{-1}$ the curves are identical. Therefore, we can choose at time $12<\,t<\,30$ MeV$^{-1}$ and sample the overlap probability for some value $\delta_V$ and interpolate the obtained data with a cosine function $A\,\cos^2(\delta_V-B)$. The maximum value of the function corresponds the exact phase shift $\delta_L$.  
	
\begin{figure}[th]
		
		\includegraphics[width=\columnwidth]{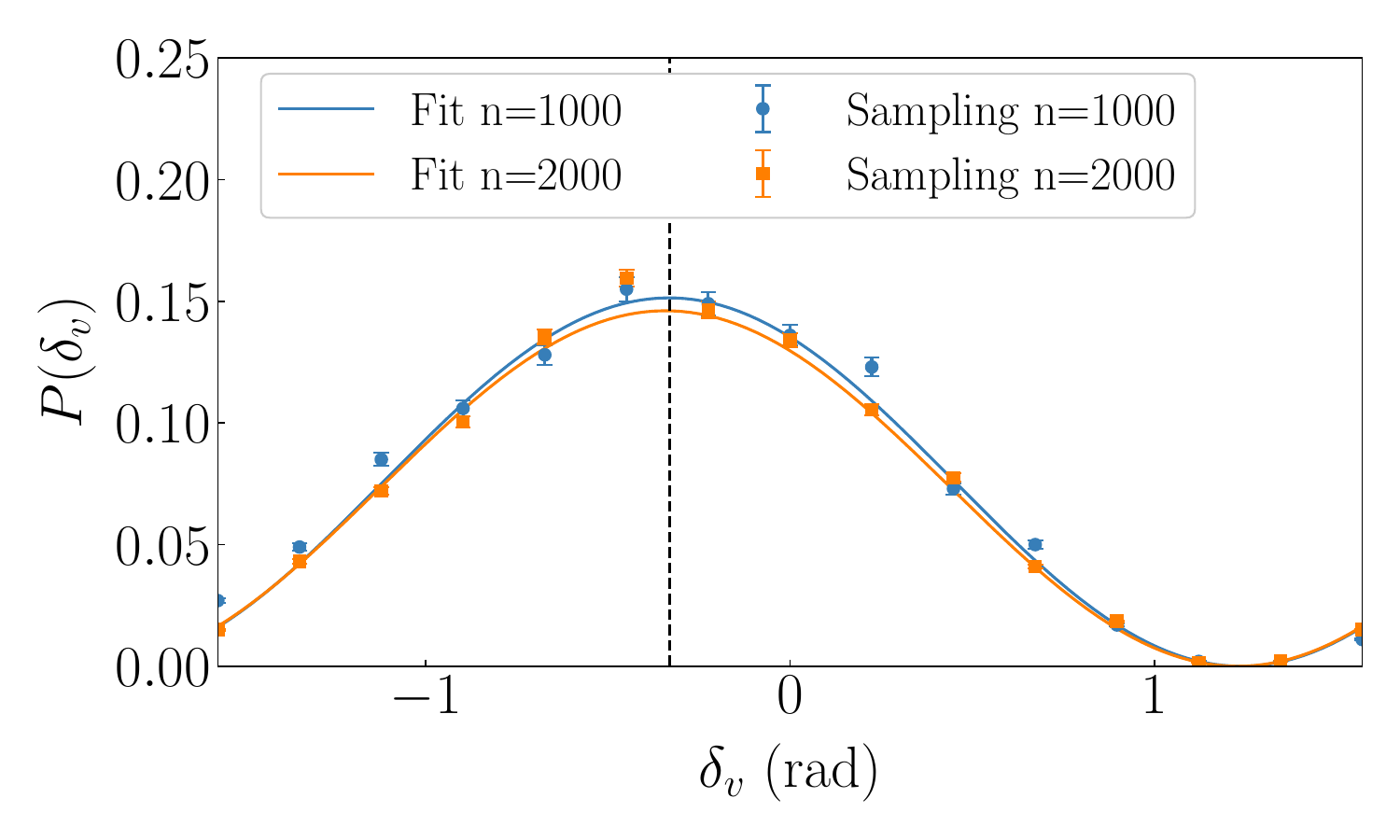}
		\caption{Sampled probability as a function of $\delta_V$ for different shots at time $t=20$ MeV$^{-1}$. The points correspond to the obtained results, and lines indicate the obtained fit. The vertical dashed line shows the exact phase shift.}
		\label{fig:variational_fit}
	\end{figure}

	We apply this idea considering a noiseless quantum processor. We sample the overlap probability at time $t=20$ MeV$^{-1}$ for different numbers of shots. We obtain the results shown in Fig.~\ref{fig:variational_fit}. The lines represent the fitting results whose phase parameters are reported in Table~\ref{tab:variational_fit}. The obtained phase shifts are compatible inside two sigma with the exact one.
	
	The proposed algorithm can be implemented in two steps in order to obtain the phase shift of a generic unknown quantum system:
	\begin{enumerate}
		\item Apply the TEPS method to obtain the time interval when the overlap probability becomes flat.
		\item Choose a time $t$ inside this interval. Compute the overlap probability $P(\delta_V)$ as a function of some points $\delta_V$. Then, interpolate them with a cosine square function and extract the phase shift value. 
	\end{enumerate}

	\begin{table}[th]
		\caption{Results of the fit of Fig.~\ref{fig:variational_fit}.}
		\begin{tabular}{|c|c|c|}
			\hline
			$N_{shots}$ & Fitted $B$ & Exact $\delta_L$ \\
			\hline    
			1000 & -0.336(8)&-0.330\\
			2000 & -0.33(1)&-0.33\\
			\hline
		\end{tabular}
		
		\label{tab:variational_fit}
	\end{table}

	\subsection{Expansion in Spherical Bessel functions}
	
	\begin{figure*}[ht]
		
		\includegraphics[width=\textwidth]{ 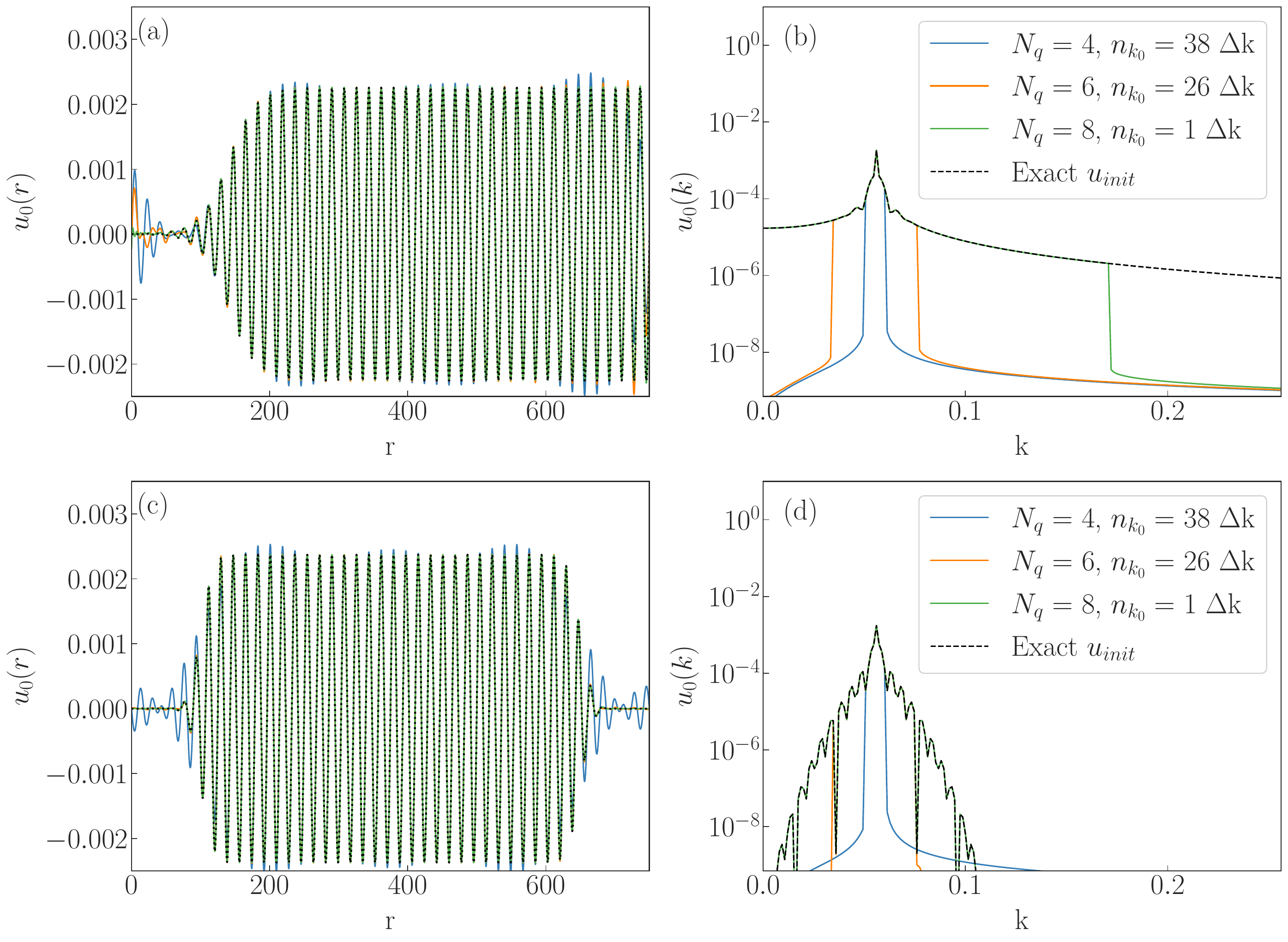}
		\caption{The obtained initial wave functions for different Bessel basis set for a truncated plane wave with momentum $k_{\rm in}=0.351$ fm$^{-1}$. Panel (a) and Panel (b) show their spatial and momentum distribution when we filter with Eq.~\eqref{eq:filter_fermi}. Instead, Panel (c) and Panel(d) shows the obtained spatial and momentum distribution when we filter with Eq.~\eqref{eq:erf_filter}.
			The legend reports the number of qubits and the lower momentum of the expansion with $\Delta k = 0.0084$.}
		\label{fig:initialbessel}
	\end{figure*}
	
	We implement the V-TEPS algorithm of Sec.~\ref{sec:variational} when we expand the Hamiltonian in the momentum basis. The basis set for each momentum is given by the first and second kind spherical Bessel functions for the non-relativistic case.
	
	As described in Sec.~\ref{sec:realtime}, we have to start from a truncated version of the plane wave.Figures \ref{fig:initialbessel}a and \ref{fig:initialbessel}b show an example for the obtained initial wave functions with $k_{\rm in}=0.351$ fm$^{-1}$ using different momentum basis sets using Eq.~\eqref{eq:filter_fermi} as a filter function. Panel (a) shows its spatial distribution, and panel (b) shows its momenta composition. In the plots, $N_q$ indicates the number of qubits, given by the double of the used momenta\footnote{For each momentum $k_i$, we have two spherical Bessel functions.}, $n_{k_0}$ the lowest momentum in the basis set, and $\Delta k=\frac{2\pi}{R_{max}}$, with $R_{max}=750$ fm. We can observe wrong behaviors of the initial wave function in Fig.~\ref{fig:initialbessel}a for $r\xrightarrow[]{}0$. 
	This is explained by the asymmetry of the filter function. Indeed, for a finite Box, we can also pick contributions from $y_l$ (for $L=0$ this is described by $y_0=\frac{\cos(k\,r)}{r}$)\footnote{We have also expanded the initial wave function only with $j_0$ functions but the obtained results are not correct.}. Expanding with higher momenta mitigates this contribution. 
	
	Therefore, our implemented filter function for the initial state and detectors is given by
	
	\begin{equation}
		f(r) = \frac{A}{2} \left( \text{erf}\left( \frac{r-r_1}{\Gamma}\right)+\text{erf}\left( \frac{r_2-r}{\Gamma}\right)  \right) \,, \label{eq:erf_filter}
	\end{equation}
	
	where $\erf(t)$ is defined as 
	\begin{equation}
		{\rm erf}(t)=\frac{2}{\sqrt{\pi}}\int_{0}^t dx\, e^{-x^2}\,,
	\end{equation}
	where $r_1$ and $r_2$ describe the region where the wave function is not $0$, and $\Gamma$ represents the width of the filter. 
	
	Figures~\ref{fig:initialbessel}(c) and~\ref{fig:initialbessel}(d) show the spatial distribution and their Fourier transform, respectively, for the initial plane with $k_{\rm in}=0.351$ fm$^{-1}$ using Eq.~\eqref{eq:erf_filter} as filter. We observe a good compatibility between the obtained initial wave function and the exact one (black dashed line). Using the momentum basis set, we get an advantage. Only a few momenta mainly contribute to the initial wave function and, hence, this state can be easily prepared on a quantum processor.

\begin{figure*}
		
		\includegraphics[width=\textwidth]{ 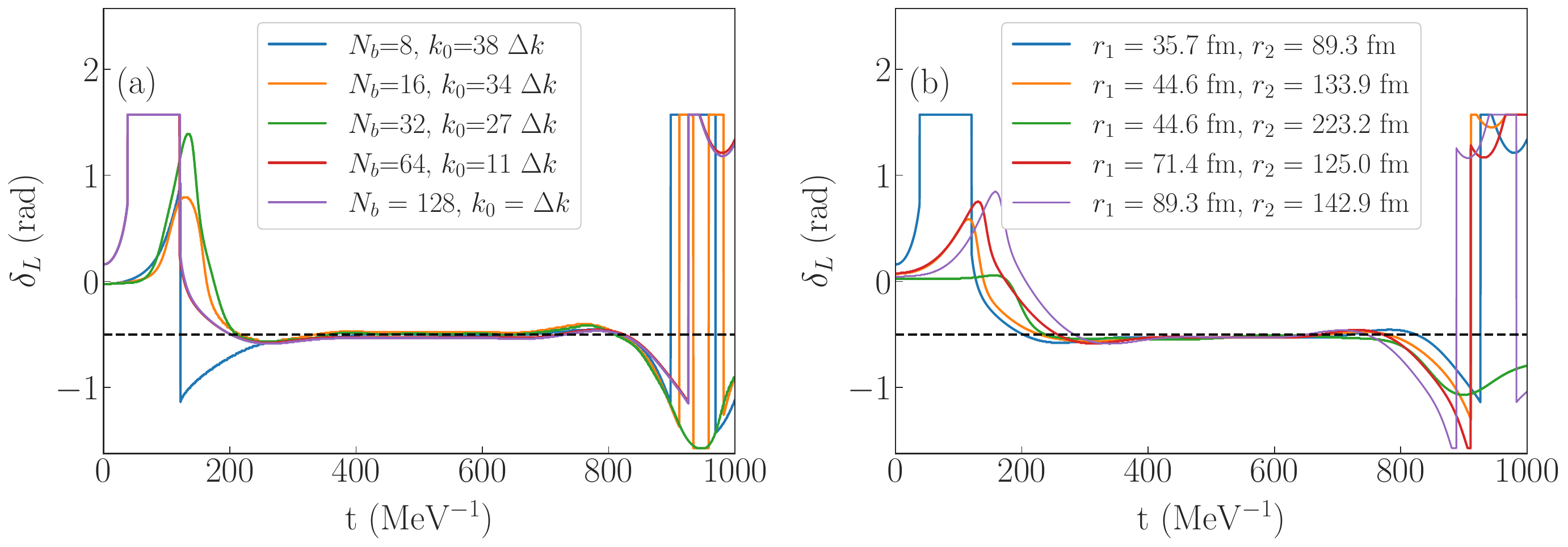}
		\caption{ In Panel (a), phase shift results for different used numbers of qubits. In the legend, $N_b$ indicates the number of momenta and $k_0$ the used initial momentum. The spacing of the phase shift grid is $\frac{\pi}{100}$, and the momentum spacing is $\Delta k = 0.0084$. Panel (b), phase shift results for different detector positions, changing the interval $[r_1,r_2]$ of the box functions.}
		\label{fig:phase_bessel_vs_qubits}
	\end{figure*}

	We implement the V-TEPS algorithm for a Gaussian potential with $V_0=1$ MeV and $\sigma=2$ fm. Figure~\ref{fig:phase_bessel_vs_qubits} shows the obtained results for a different number of expanded momenta (we used the same basis set of Fig.~\ref{fig:initialbessel}). All the different lines converge to the exact phase shift value, which is indicated by the horizontal dashed black line. The increased number of expanded momenta in the basis set shows that the phase shift values become flatter and last longer time. 
	We highlight the blue line simulation because, using four qubits, we can obtain a good estimation of the phase shift. Indeed, we implement this simulation on the IBM processors. The next section discusses this implementation in detail.
	
	Figure~\ref{fig:phase_bessel_vs_qubits}(b) shows how the phase shift results depend on different detector positions. We observe that all the lines reach the exact phase shift value. If the detector gets farther away from $r=0$, there is a longer delay in reaching the exact result interval, as shown with the green line behaviors. Moreover, the detector  for the green line is far away, and it is closer to the right infinite wall ($R_{max}=750$ fm). Therefore, the plateau interval is shorter and nosier due to boundary effects.

	V-TEPS results for different Gaussian potentials and for H-Kr Lennard-Jones potential are reported in Table~\ref{tab:bessel_phaseshift} for different initial momenta. In the table, we specify the set-up of the simulations, reporting the number of used momenta ($N_b$), the starting momentum of the basis set ($k_0$), and the momentum spacing ($\Delta k)$. These results are compatible with the exact ones within two standard deviations. 
	We can also observe that by enlarging the momentum basis set, the phase shift results converge to the exact value.

	\begin{table}[ht]
		\caption{Results of the phase shift obtained applying the V-TEPS method for the Bessel expansion. In the Table, we report the number of expanded momenta ($N_k$), the momentum spacing ($\Delta k$), and the lower expanded momentum ($k_0$).}
		
		\begin{tabular}{|c|ccc|c|c|}
			\hline
			\multicolumn{6}{|c|}{Gaussian $V_0=1$ MeV $\sigma=2$ fm}\\
			\hline
			$k_{\rm in}$ (fm$^{-1}$)& $N_k$ & $\Delta k$ & $k_0$ & Result &Exact\\
			\hline
			0.351 &8 & 0.0084 & 0.315 & -0.48(2) &  -0.50\\
			0.351&32 & 0.0084 & 0.224 & -0.492(12) &  -0.499\\
			0.351 &128 & 0.0084 & 0.0084 & -0.53(1) &-0.50\\
			
			0.545& 128 & 0.0084 & 0.0084  & -0.723(1) &  -0.704\\
			\hline
			\multicolumn{6}{|c|}{Gaussian $V_0=2$ MeV $\sigma=4$ fm}\\
			\hline
			$k_{\rm in}$ (fm$^{-1}$)& $N_k$ & $\Delta k$ & $k_0$ & Result &Exact\\
			\hline
			0.415& 128 & 0.0084 & 0.0084 & 1.00(1) &1.00\\
			0.662& 128 & 0.0084 & 0.0084  & 0.13(1) &  0.11\\
			1.014  & 256 & 0.0084 & 0.0084 & -0.457(12)&-0.47\\
			
			\hline
			\multicolumn{6}{|c|}{Lennard Jones H-Kr}\\
			\hline
			$k_{\rm in}$ (\r{A}$^{-1}$)& $N_k$ & $\Delta k$ & $k_0$ & Result &Exact\\
			\hline

			0.408 & 256 &0.013&0.0157&-1.45(5) &-1.33 \\
			0.408 & 512 &0.013&0.013&-1.33(3) &-1.33 \\
			0.537&600 &0.013 & 0.0157 &1.09(4)& 1.10
			\\
			1.06&600 &0.013 & 0.0157 &-1.44(3)& -1.44\\
			\hline
		\end{tabular}
		
		\label{tab:bessel_phaseshift}
	\end{table}

	\section{Results from IBM device \label{sec:results_quantum_hardware}}
	
	We simulate the four qubit case of Fig.~\ref{fig:phase_bessel_vs_qubits} on the free-access IBM quantum processors~\cite{ibm,Qiskit} to diagnose the contribution of the quantum noise on the proposed V-TEPS algorithm. We only changed the filter function for the initial state, using the one in Eq.~\eqref{eq:filter_fermi} because in this case it gives cleaner data. Its implementation was done in two steps. First, we evaluate the time dependence of the overlap probability with the free plane wave (setting $\delta_V=0$ ) to search when the probability is flat.
	Second, we fix the time $t$ inside this interval, and we compute the overlap probability as a function of the detector phase $\delta_{V}$. The phase shift of the scattering process is obtained from the maximum position.

	The coherence time of noisy intermediate-scale quantum era superconducting devices is too short to study the application of Trotter decomposition. For this reason, we implement the full real-time evolution with a single-time step.
	
	We map the state in the computational basis with Bessel functions. Applying the python \textit{OpenQL} package~\cite{openql} for compiling the quantum circuit, this requires 196 CNOT gates. With such a large number of CNOT gates, we reach the decoherence of the quantum processors, where the probability falls to the decoherence line (1/16 in this case).
	To overcome this problem, we map the computational basis into the eigenstate of the Hamiltonian. This choice of mapping shrinks the depth of the quantum circuit (63 CNOT gates in total) because the real-time evolution operator is diagonal. 
	
\begin{figure}
		
		\includegraphics[width=\columnwidth]{ 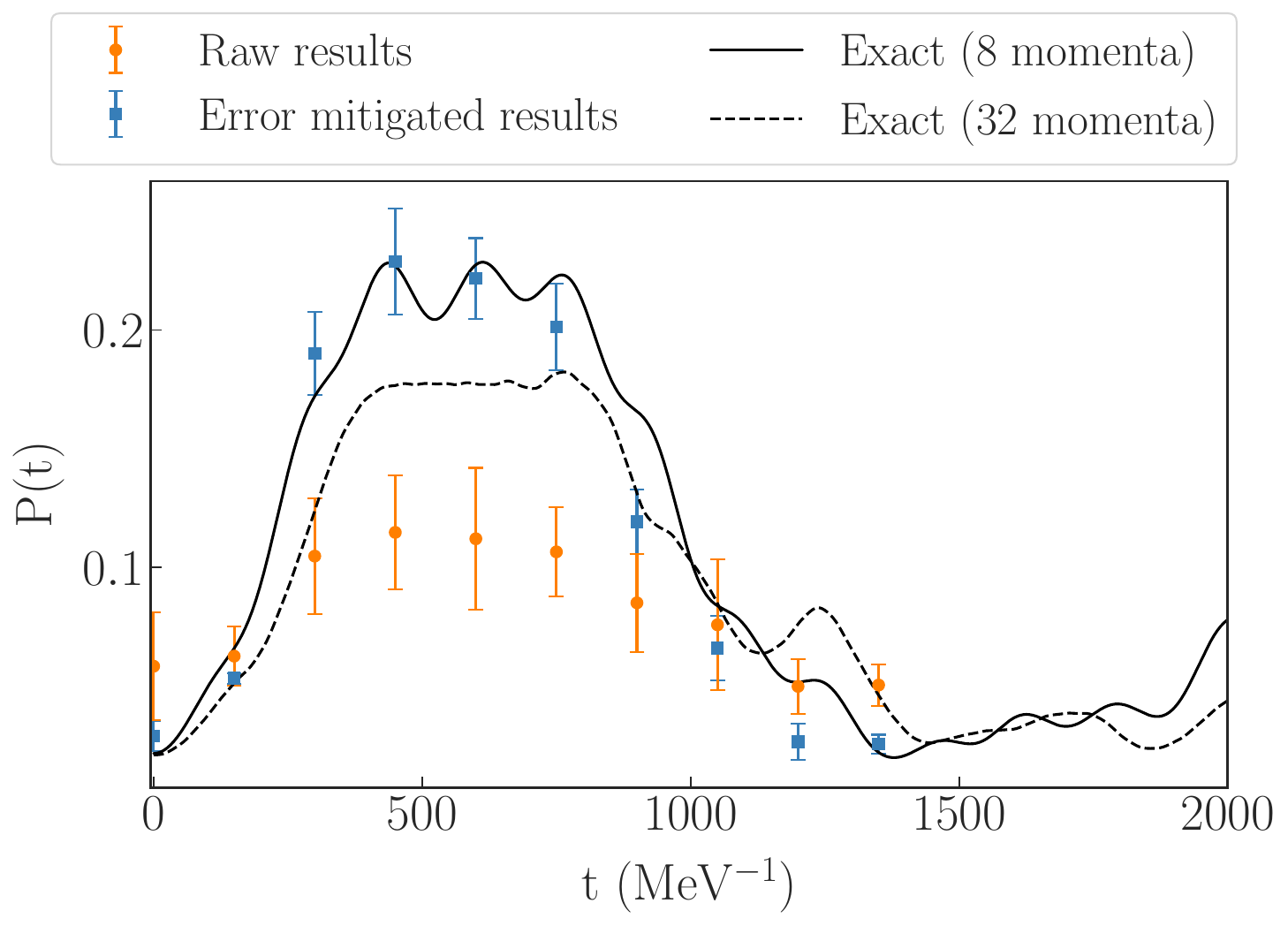}
		\caption{Obtained detector probability as a function of time. Circles and squares indicates the raw and error-mitigated results obtained from the \texttt{ibm\textunderscore perth}, respectively. The solid black line represents the noiseless result. The dashed line reports the exact case when we expand with more momenta (32).} \label{fig:real_time_bessel_qc}
	\end{figure}

	Implementing this simple quantum circuit (shown in App.~\ref{app:ibm_quantum_circuit}) on the \texttt{ibm\textunderscore perth} device, we obtain the orange circles and blue squares shown in Fig.~\ref{fig:real_time_bessel_qc}. In both results, we implement the readout error correction and Pauli twirling~\cite{Pauli_twirling}, where for each of the points we run eight different Pauli-twirled quantum circuits. The blue circles, were obtained by also applying the decoherence renormalization (DR) method~\cite{Decoherence_ren1,Decoherence_ren2,Decoherence_renorm} that mitigates decoherence effect. Details of DR algorithm can be found in App.~\ref{app:DR}.
	
	We observe that the probability results in Fig.~\ref{fig:real_time_bessel_qc} have a plateau between $t=450$ MeV$^{-1}$ and $t=750$ MeV$^{-1}$), providing a good interval to apply the second step of the V-TEPS algorithm. Moreover, the blue squares become closer to the analytical evolution, that is shown with the black line. We also report the exact case when we using 32 momenta (dashed line), showing that the probability gets flatter when enlarging the number of momenta. 
	
\begin{figure}
		
		\includegraphics[width=\columnwidth]{ 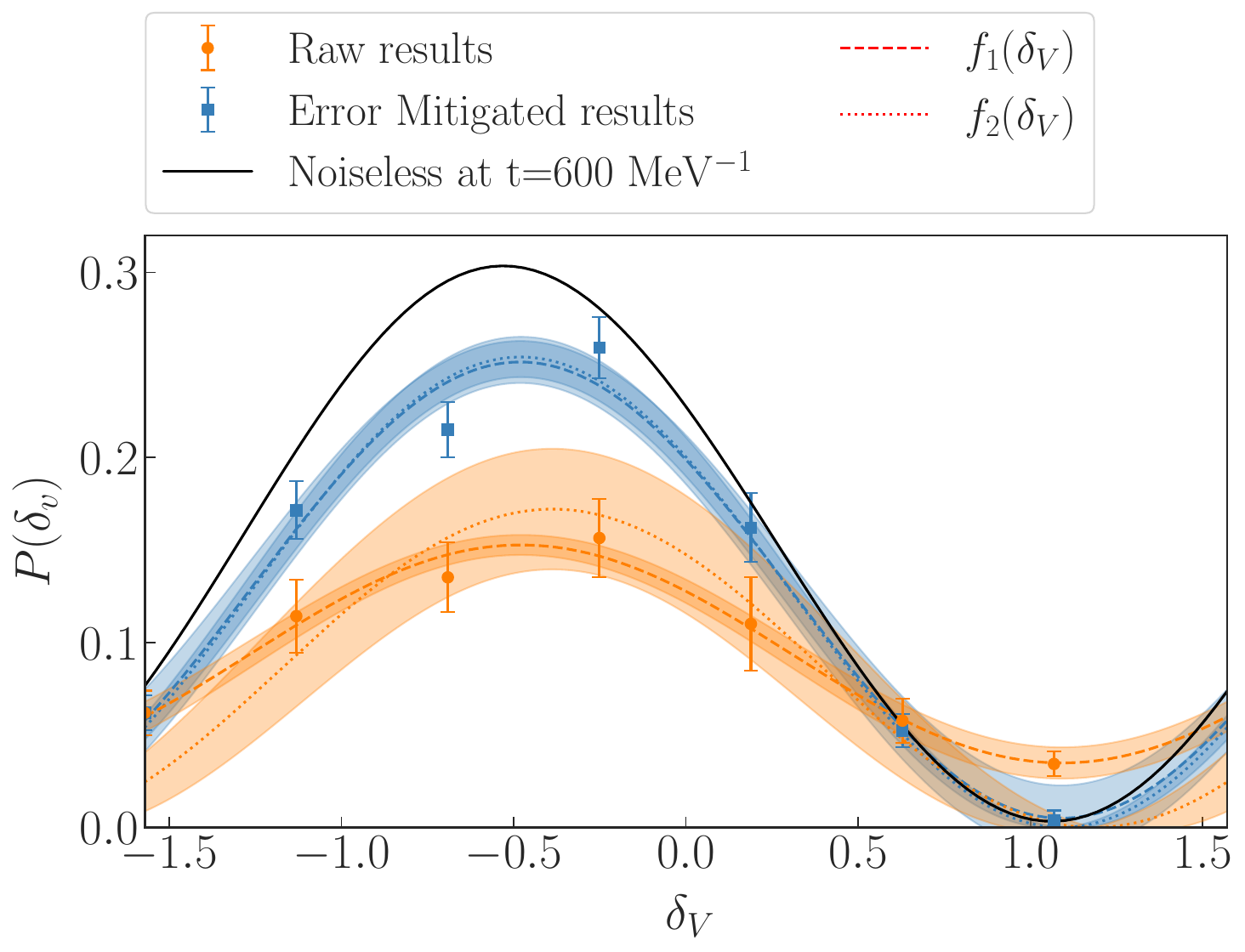}
		\caption{Obtained probability as a function of $\delta_v$. Circles and squares indicates the raw and error-mitigated results obtained from the \texttt{ibm\textunderscore brisbane}, respectively. The obtained fitted curves are shown with dashed lines. Black solid line corresponds to the noiseless values.}
		\label{fig:delta_V_bessel_qc}
	\end{figure}

	Inside the plateau, we choose the time $t=600$ MeV$^{-1}$ to implement the second step of the variational method. Figure~\ref{fig:delta_V_bessel_qc} shows the obtained results from the \texttt{ibm\textunderscore brisbane} device, where we obtain the overlap probability as a function of the shifted phase $\delta_V$ in the detector plane wave. Circles and squares represent the results with the same legend of Fig.~\ref{fig:real_time_bessel_qc}. The classical noiseless result is indicated with the solid black line. We interpolate the two set of points with two functions, 
	\begin{equation}
		f_1(\delta_V)=c_1+a_1\,\cos^2(\delta_V+b_1)
	\end{equation} 
	and
	\begin{equation}
		f_2(\delta_V)=a_2\,\cos^2(\delta_V+b_2)\,,
	\end{equation}
	where in $f_1$ we use a constant $c_1$ to describe the effect of decoherence line.
	
	The two dashed and dotted lines in Fig.~\ref{fig:delta_V_bessel_qc} represent the interpolated results where the obtained fitting parameters are reported in Tab.~\ref{tab:fitting_values}.  The color areas represent the one-sigma uncertainties of interpolations. If we compute from the fit curve the position of the decoherence line, we obtain $D_L=0.094(6)$, which is close to the expected value of $1/16=0.0625$.
	
	Table~\ref{tab:fitting_phase_shift_qc} shows maximum position of each fits. All the results are two sigma compatible with the exact phase shift. We observe that the effect of noise is to level the probability down to the decoherence line (given by $\frac{1}{2^n}$ with $n$ number of qubits). If we do not reach the full decoherence time, the proposed quantum algorithm is capable of estimating the correct phase shift because the noise contribution does not change the maximum position.
	
	\begin{table}[ht]
		\caption{Fitting results of Fig.~\ref{fig:delta_V_bessel_qc} for $f_1$ and $f_2$}
		\begin{tabular}{|c|ccc|}
			\hline
			&$a_1$&$b_1$&$c_1$\\
			\hline
			Raw $f_1$&-0.118(7) & 1.09(3)  &0.153(5)\\
			Err. mitigated $f_1$ & 
			-0.247(14) & 1.09(3) &  0.252(11)\\
			\hline
			&$a_2$&$b_2$&\\
			\hline
			Err. mitigated $f_2$ & 0.254(11) &-0.48(3) &\\
			Raw $f_2$& 0.17(3) &-0.39(12)&\\
			\hline
		\end{tabular} 
		\label{tab:fitting_values}
	\end{table}

	\begin{table}[ht]
		\caption{Phase shift results obtained from the maximum position of the interpolation curves. }
		\begin{tabular}{|c|c|}
			\hline
			& Phase shift (rad)\\
			\hline
			Exacts (from Numerov) & -0.500 \\
			Noiseless at $t=600$ MeV$^{-1}$ & -0.49(3)\\ 
			\hline
			Fitting $f_1(\delta_V)$ mitigated results & -0.48(3)\\ 
			Fitting $f_2(\delta_V)$ mitigated results & -0.48(3)\\ 
			Fitting $f_1(\delta_V)$ raw results & -0.48(3)\\ 
			Fitting $f_2(\delta_V)$ raw results & -0.39(12)\\ 
			\hline
		\end{tabular}
		
		\label{tab:fitting_phase_shift_qc}
	\end{table}

	\section{Conclusions}
	\label{sec:conclusions}
	
	We present a quantum algorithm that evaluates the phase shift for non-relativistic elastic scattering processes for a generic Hamiltonian.
	
	We start from a truncated plane wave, which goes to zero close to the potential. Then, we evolve the wave function in real-time. The phase shift is computed from the overlap probability with a target plane wave (detector) that is localized in a small region far away from the interaction. After some time, the phase shift evolution converges to a plateau whose value is the exact phase shift. Numerical classical results are reported and discussed for different Hamiltonians.
	
	We also present an improved version of the proposed algorithm with a variational approach, shifting the phase of the detector plane wave with a fictitious parameter $\delta_V$. The exact phase shift is recovered by the value of $\delta_V$ that maximizes the overlap probability. This variational approach improves the phase shift results, reducing their variance. 
	
	Implementing the proposed quantum algorithm for general complex scattering processes, one may apply the real-time evolution algorithm to scan the time interval to find the plateaus in the phase shift. Then, fixing a time $t$ where the phase shift is constant, one should implement the variational method for evaluating precisely the phase shift. For the variational approach, numerical results and discussions are reported for different Hamiltonians and basis sets (spatial lattice, momentum basis sets).
	
	All the classical simulations show the reliability of the proposed quantum algorithms. We also presented quantum simulations implemented on IBM quantum processors, where we expanded the Hamiltonian in a smaller momentum basis set. We demonstrate that we can obtain the exact phase shift result by interpolating a few values. Indeed, searching just the position of the maximum, the noise source from quantum processors does not affect the phase shift results because it only flattens the probability, not the maximum position. 
	
	This algorithm is guided in two main different processes, a state preparation and real-time evolution. In the momentum expansion, the state preparation for the initial plane wave and detector can be done with a shallow local quantum circuit. Indeed, only the momenta close to the initial value of the plane wave $k_{\rm in}$ contributes to the expansion of the state. Therefore,
	the scaling of the proposed quantum algorithm is mainly due to the scaling of the real-time evolution. 
	
	The presented quantum algorithms are a step forward in simulating non-relativistic scattering processes with quantum computers. Indeed, this algorithm can be implemented to compute phase shifts and cross sections for elastic scattering both in single and multi-channel problems. This variational algorithm may be extended to the inelastic scattering case. In practice, inelastic scattering can be treated, in approximation, by including numerous reaction outcomes (e.g. channels, states, reaction products, etc.) in the simulation while only measuring the outcome of interest.
	Including all, the proposed quantum algorithm can be implemented to obtain the result of reactions because the variational approach can be implemented in a generic scattering experiment, where the momentum is coupled to recover the correct phase shifts.

	\section{Acknowledgments}

	We thank the IQuS group for useful discussions. In particular, in the IQuS group, we are grateful to Marc Illa Subi\~{n}\`{a}, Saurabh Vasant Kadam, Xiaojun Yao, Anthony Ciavarella, Rolland Farrell and Martin Savage for the useful discussions. Also, we thank Marc Illa and Rolland Farrell for the explanation of the decoherence renormalization method.

	This work was supported in part by the U.S. Department of Energy, Office of Science, Office of Nuclear Physics, InQubator for Quantum Simulation (IQuS) (\url{https://iqus.uw.edu}) under Award Number DOE (NP) Award DE-SC0020970 via the program on Quantum Horizons: QIS Research and Innovation for Nuclear Science.
	This work
	was prepared in part by LLNL under Contract No. DE-
	AC52-07NA27344 with support from the U.S. Department of Energy, Office
	of Science, Office of Nuclear Physics (under Work Proposal
	No. SCW1730).
	
	This work was enabled by the use of
	advanced computational, storage, and networking infrastructure provided by the Hyak supercomputer system at
	the University of Washington (\url{https://hyak.uw.edu/}).
	
	We acknowledge the IBM Quantum services for this work. The views expressed are those of the authors, and do not reflect the official policy or position of IBM or the IBM Quantum team.

	\bibliography{reference}
	
	\appendix

	\section{Overview of used Hamiltonians \label{app:Hamiltonian}}
	This section illustrates the used potentials. 
	The Gaussian potential is given by,
	\begin{equation}
		V(r)= V_0 e^{-\frac{r^2}{\sigma^2}} \,,   
	\end{equation}
	where $V_0=1$ MeV and $\sigma=2$ fm. We set the reduced mass $\mu$ of the system such that $\frac{\hbar^2}{2\mu}=1$ MeV fm${}^2$.
	
	The Lennard-Jones potential is given by  
	\begin{equation}
		V(r)= 4 \epsilon \left[\left(\frac{\sigma}{r}\right)^{12}-\left(\frac{\sigma}{r}\right)^{6}\right] \,, \label{eq:Lennard_jones}
	\end{equation}
	where $\sigma\,=\, 3.57$ \r{A} and $\epsilon\,=\,5.9$ meV.  These values are obtained from the Hydrogen-Krypton scattering of Ref.~\cite{Lennard_Jones_potential}. In our simulation, we modify Eq.~\eqref{eq:Lennard_jones} as it follows,
	\begin{equation}
		V(r)=\left\{ \begin{matrix}
			4 \epsilon\,\left[\left(\frac{\sigma}{r}\right)^{12}-\left(\frac{\sigma}{r}\right)^{6}\right] &\text{  for }r>r_C \,,\\
			4 \epsilon\,\left[\left(\frac{\sigma}{r_C}\right)^{12}-\left(\frac{\sigma}{r_C}\right)^{6}\right] &\text{  for } r<r_C\,,
		\end{matrix}\right.  \end{equation}
	where $r_C=0.4\,\sigma$ indicates a spatial cutoff in order to avoid numerical issues for the hard-core potential. The used parameters can be found in Ref.~\cite{Lennard_Jones_potential}
	
	\section{Phase shift dependence on real-time parameters \label{app:parameters_algorithm}}
	
	This section reports how the TEPS parameters contribute to the phase shift results for the Gaussian potential with $V_0=2$ MeV and $\sigma=4$ fm with an initial momentum $k_{\rm in}=1.9107$ fm$^{-1}$ using the spatial finite box. In particular, these parameters are given by the number of lattice points $R_{\rm p}$, lattice spacing $a$, filter parameters $\Gamma$ and $r_0$, and position and width of the detector, $r_1$ and $r_2-r_1$. In order to find how errors correlate to these parameters, for each simulation we only vary one; the others are fixed. The standard parameter values are $R_{\rm p}=6000$, $a=0.02$ fm, $r_0=26$ fm, $\Gamma=2$ fm, $r_1=16.44$ fm,  $r_2-r_1= 23.02$ fm.
	
\begin{figure*}[h]
		
		\subfloat[Phase shift results for different lattice points $R_{\rm p}=\frac{R_{max}}{a}$.]{\includegraphics[width=\columnwidth]{  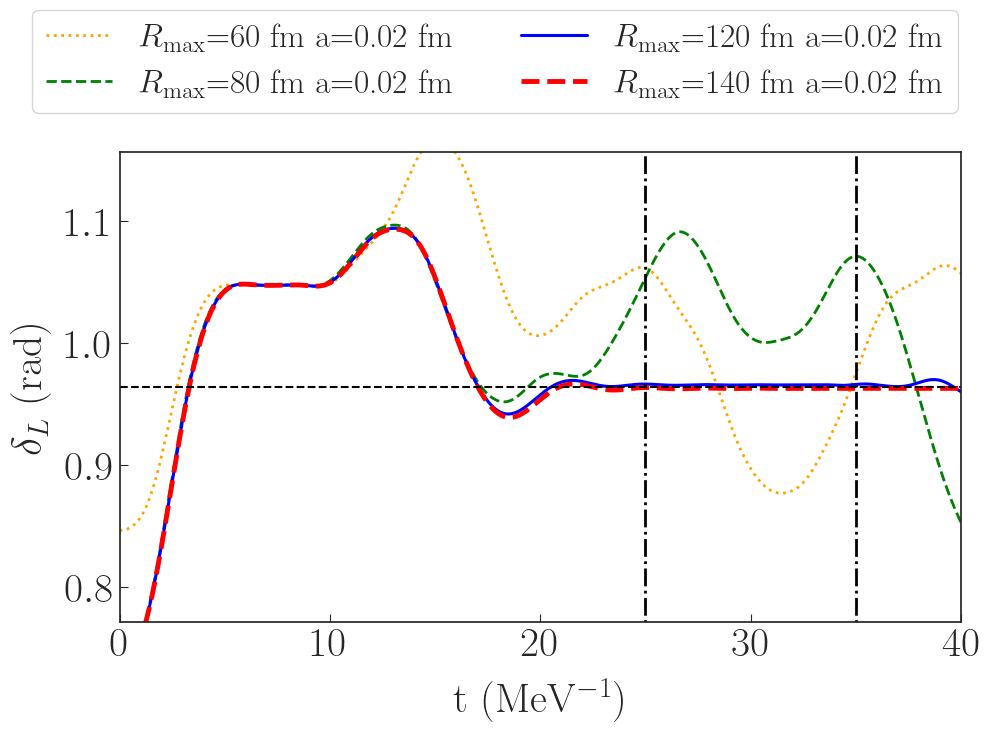}\label{fig:phaseshift_L} }\subfloat[Phase shift results for different lattice spacing $a$.]{\includegraphics[width=\columnwidth]{  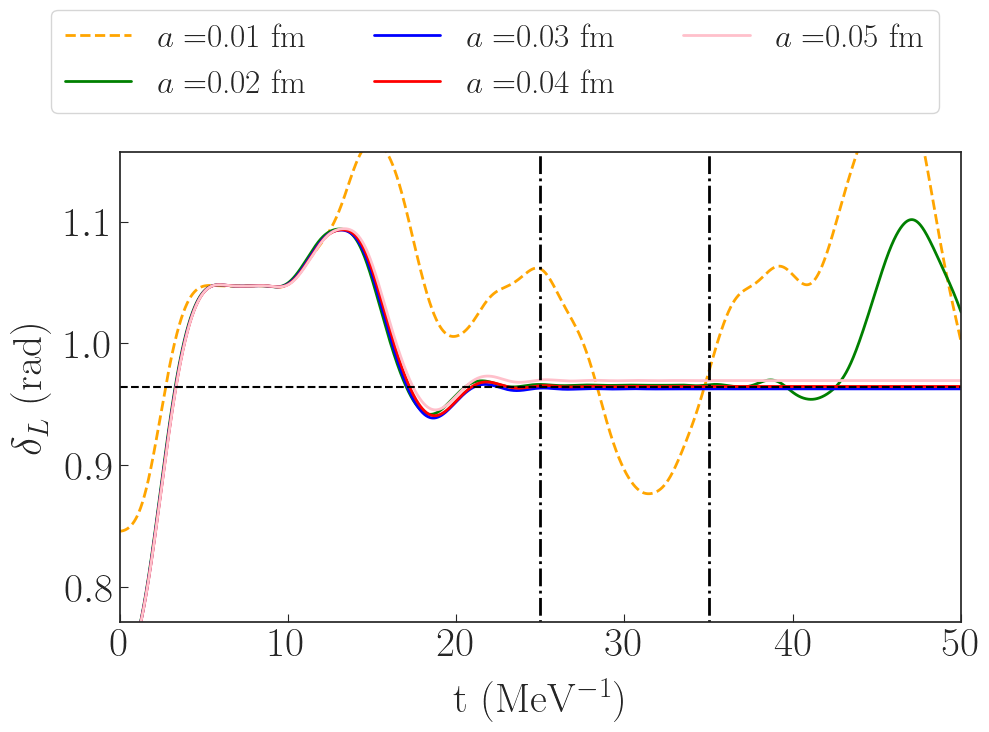}
			\label{fig:phaseshift_a}}\\
		
		\subfloat[Phase shift results for different lattice spacing $a$ and fixing the volume $a\,R=120$ fm. ]{\includegraphics[width=\columnwidth]{  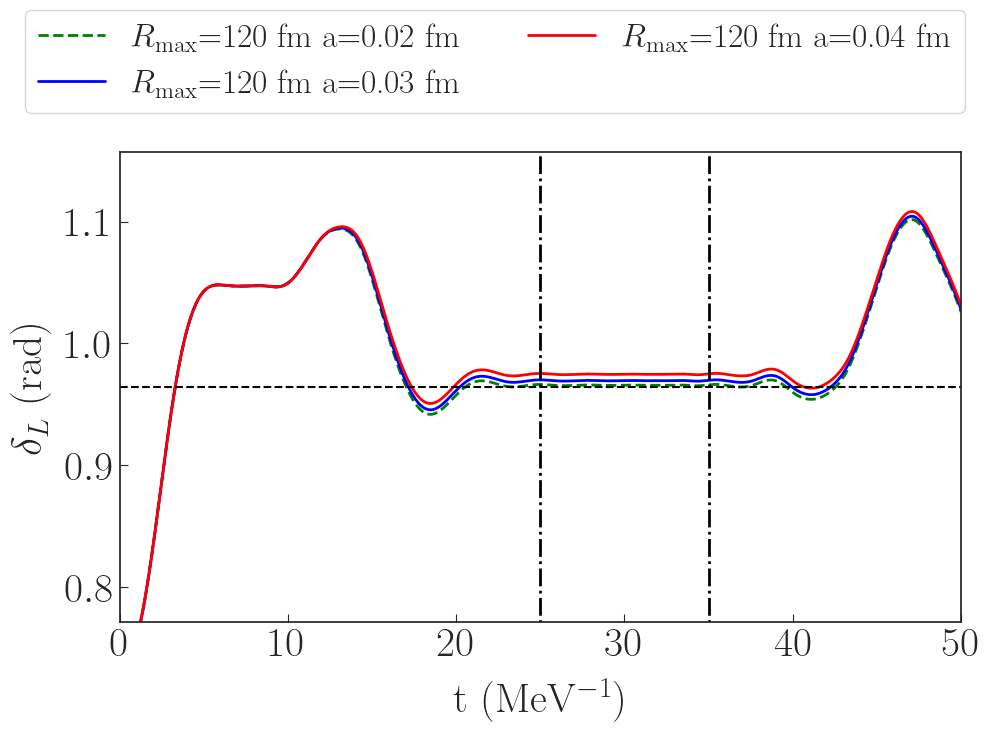}\label{fig:phaseshift_La}}
		\caption{Phase shift results for different lattice parameters .The horizontal dashed line indicates the absolute exact phase shift value $|\delta^{ex}_L|=0.965$.}
	\end{figure*}

	We found that significant errors comes from finite volume and discretization effects ($R_{\rm p}$ and $a$). Figure~\ref{fig:phaseshift_L} shows how  the phase shift depends on the $R_{\rm p}$ value. We can observe that increasing $R_{\rm p}$, the time where the phase shift is constant gets longer. Instead, Fig.~\ref{fig:phaseshift_a} illustrates the obtained phase shift results when we vary $a$. We observe the same behavior in the case of the variation of $R_{\rm p}$. However, we start to observe some errors due to the discretization. Moreover, if the lattice is small, we obtain a weird behavior of the phase shift results, as in the case of $a=0.01$. Figure~\ref{fig:phaseshift_La} shows the dependence of the phase shift on $a$ value when we fix the volume $R_{\rm p}\,a=120$ fm. We can notice that the time when the phase shift is constant is the same, but we start to have some errors when we have fewer lattice points or bigger lattice spacing.

\begin{figure*}
		
		\subfloat[Phase shift results for different position of the initial filter $r_0$. The used filter is shown in Eq.~\eqref{eq:filter_fermi}.]{\includegraphics[width=\columnwidth]{  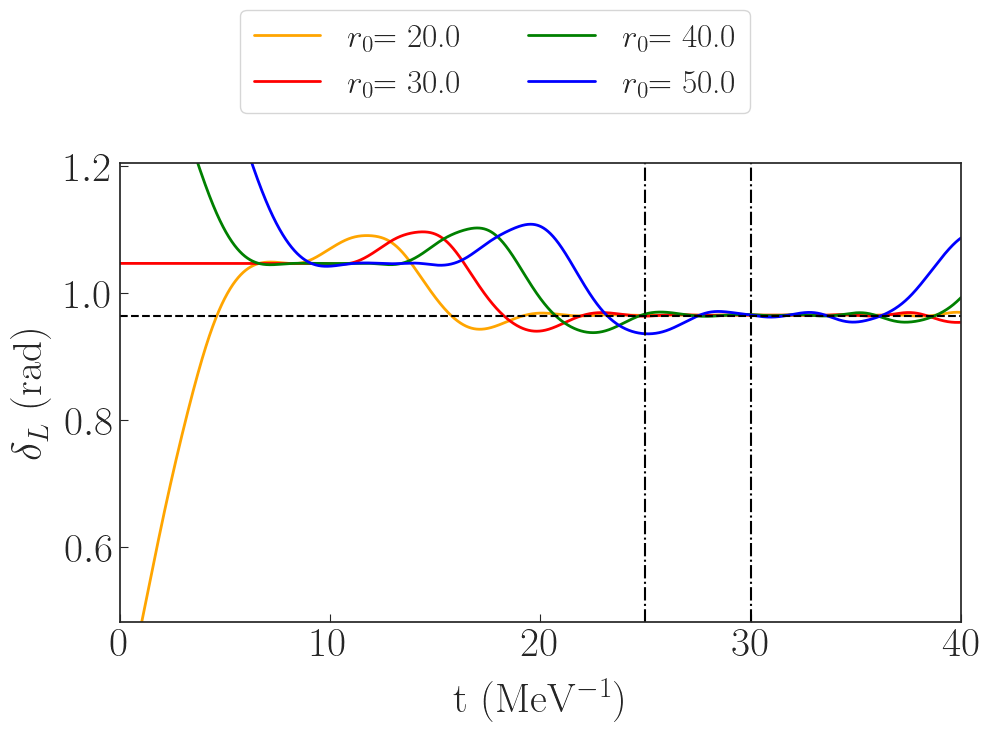}\label{fig:phaseshift_initialr0}}\quad\subfloat[Phase shift results for different position of the width filter $\Gamma$.]{\includegraphics[width=\columnwidth]{  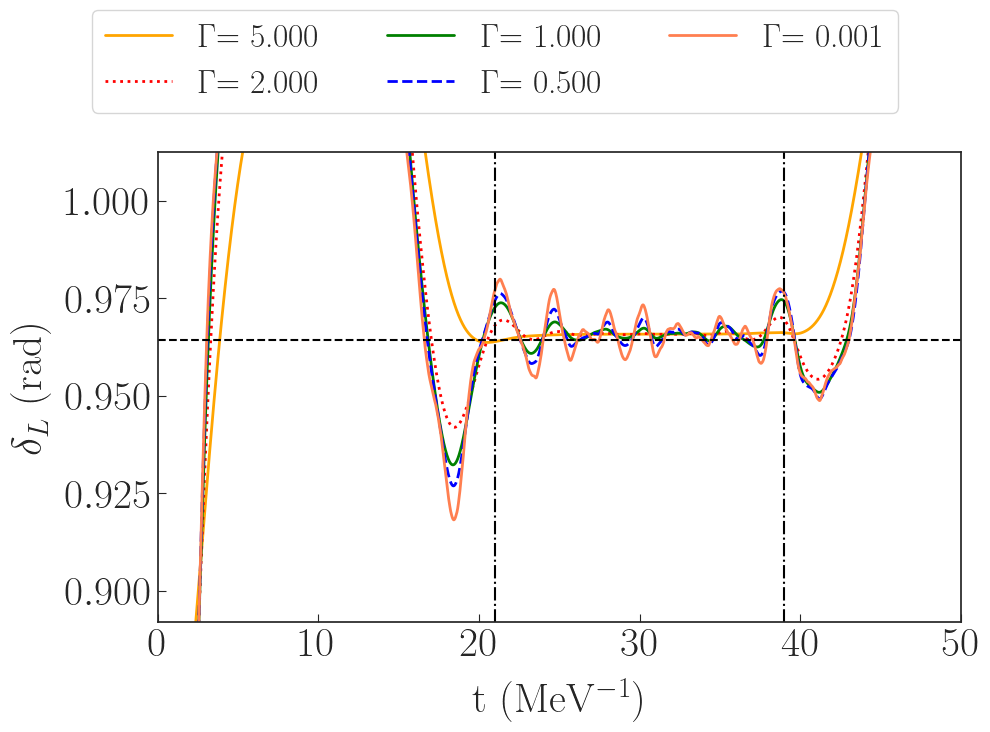} \label{fig:phaseshift_initial_sigma}}\\
		\subfloat[Phase shift results for different position of detector $r_1$. ]{\includegraphics[width=\columnwidth]{  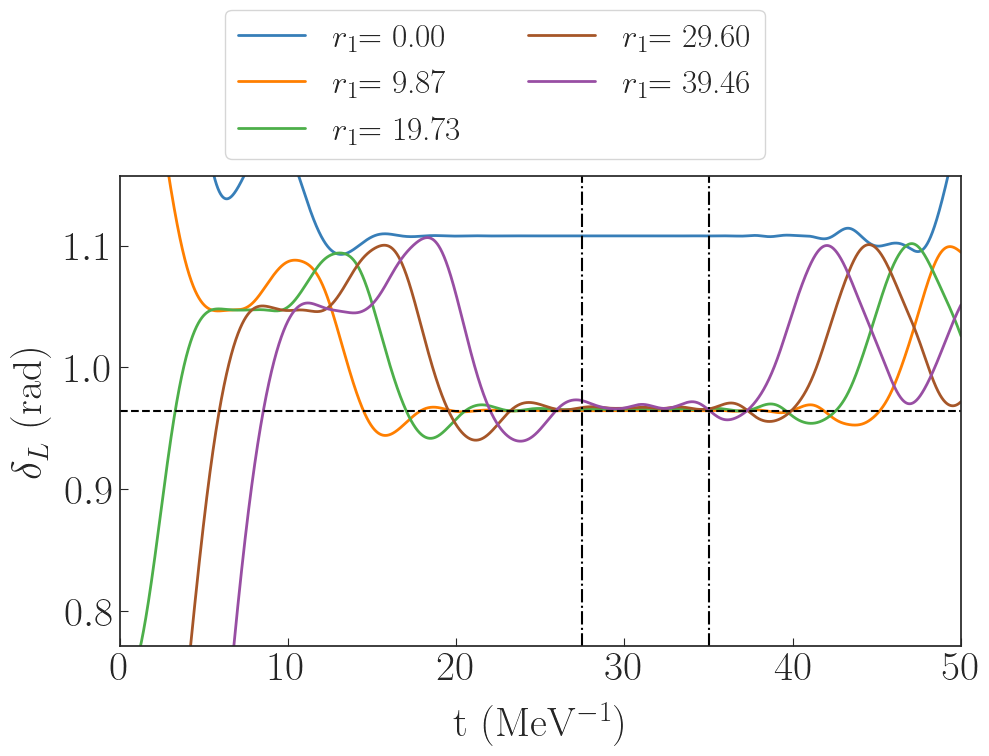}\label{fig:phaseshift_initwindow}}\quad\subfloat[Phase shift results for different position of the area of detector $r_2-r_1$.]{\includegraphics[width=\columnwidth]{  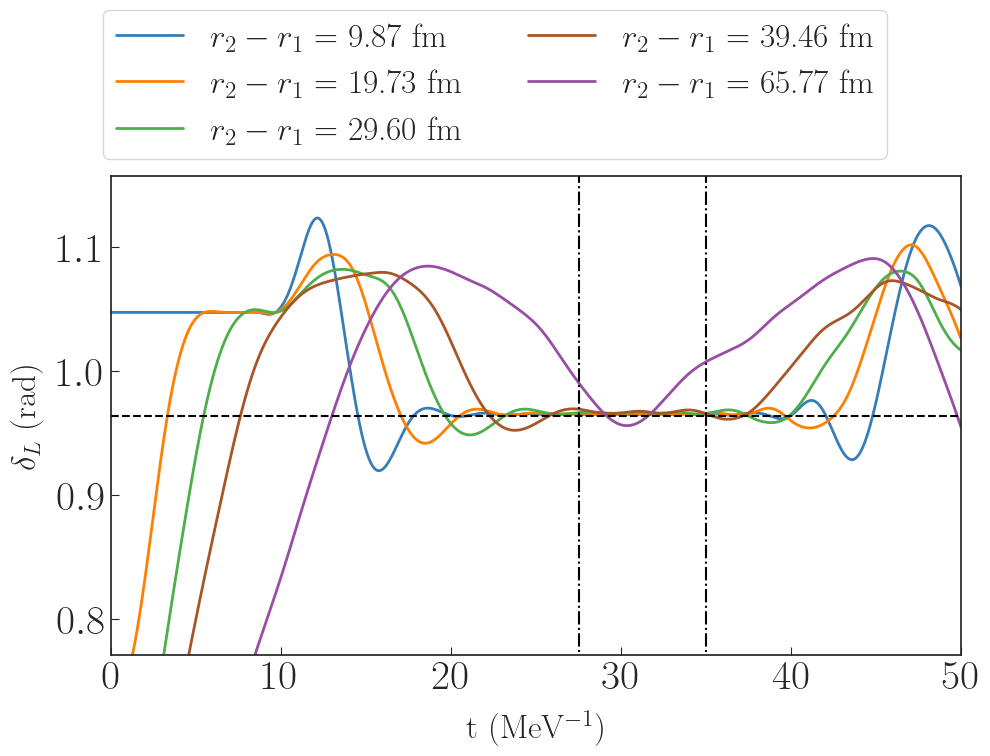}\label{fig:phaseshift_widthwindo}}\\
		
		\caption{Phase shift results when we change detector position and initial wave function. The horizontal dashed line indicates the absolute exact phase shift value $|\delta^{ex}_L|=0.965$.}
	\end{figure*}

	Figure~\ref{fig:phaseshift_initialr0} shows the phase shift dependence on the $r_0$ parameter. We can observe that our results are essentially independent of $r_0$. Instead, Fig.~\ref{fig:phaseshift_initial_sigma} shows the dependence on the $\Gamma$ parameter. The $\Gamma$ parameter is connected to the momentum truncation. For different $\Gamma$ values, we observe the variance growth in the phase shift results lowering $\Gamma$ because we start picking undesired momenta.

	The last parameters to be discussed are the detector parameters, $r_1$ and $r_2-r_1$.  Figure~\ref{fig:phaseshift_initwindow} shows that the obtained phase shift is independent of the initial position of the detector if we set it far away from the interaction zone. The only effect is just a time delay of the plateaus.  Figure~\ref{fig:phaseshift_widthwindo} reports the dependence of the phase shift from the width of the window. We obtain that the phase shift is independent from the width up to a certain value (around $35$ fm). Indeed, this value is $1/3$ of the full box, therefore, we pick the contribution from the propagation of the first part of the filter and boundary errors. However, one has to consider that the success probability depends linearly from the detector width, and the overlap probability is more sensitive to the different amplitude of the evolved wave function.

	We can conclude that for the lattice expansion, a good variance of the method presented in Sec.~\ref{sec:realtime} might be $0.03$, counting the discretization effects. However, in many cases, looking only at the phase shift results, the variance of the phase shift results is very small.\\
	Table~\ref{tab:discretization_results} reports the
	obtained results for the phase shifts.
	
	\begin{table*}[t]
		\caption{The raw obtained phase shift results varying the TEPS parameters using a spatial finite box. The phase shifts are computed for a Gaussian potential with $V_0=2$ MeV and $\sigma=4$ fm with a initial momentum $k_{\rm in}=1.9107$ fm$^{-1}$. The TEPS parameters are lattice spacing, number of lattice points, filter parameters and position and width of detector. Each column reports the obtained phase shift changing the parameter and the other are kept fixed. In the column $Ra$ we vary both $R$ and $a$ but fixing $Ra=120$. The standard parameters are $\{R=6000,\,a=0.02,\,r_0=26.,\, \Gamma=2.,\, r_1=16.44,\, r_2-r_1=23.02\}$. The exact phase shift is $\|\delta_L^{ex}\|=0.964$.   }
		
		\begin{tabular}{|cc|cc|cc|cc|cc|cc|cc|}
			\hline
			$R_{\rm p}\, a$ fixed ($a$) & $\delta_L$ & $R_{\rm p}$ &$\delta_L$ &$a$ &$\delta_L$ &$\Gamma$ & $\delta_L$ & $r_0$ & $\delta_L$ & $r_1$ & $\delta_L$& $r_2-r_1$ & $\delta_L$\\
			\hline
			0.02 & 0.9658(2) &4000 & 1.04(3)&0.01& 0.95(6) &5 & 0.9657(2) & 20 & 0.9658(1)& 0.0 & 1.1079(2)  & 9.87 & 0.9651(1)\\
			0.03 &0.9695(2)&6000 & 0.9651(3)& 0.02&0.9658(2) &2 & 0.9657(2) &30 &0.9657(5)& 9.87 &  0.9646(5) & 19.73& 0.9657(2)\\
			0.04 & 0.9748(2)&7000&  0.9652(3)& 0.03&0.9628(2)&1 & 0.9656(8) & 40 &0.9668(2)& 19.73 & 0.9657(2) &29.60 &0.9663(4)\\
			& && & 0.04&0.9648(2)&0.5& 0.965(2) & 50  & 0.96(1)& 29.60 & 0.9668(6) &39.46 & 0.9671(9)\\
			& && &0.05&0.9696(2)&0.001& 0.965(4) &  &  & 39.46 & 0.968(2) &65.77 &  0.98(2)\\
			\hline
		\end{tabular}

		\label{tab:discretization_results}
	\end{table*}
	
	\section{Proof of the variational algorithm \label{app:variational_proof}}
	
	As we discussed in the main text, the original TEPS algorithm is sensitive to the initial set-up of the simulations. In particular, when we have unstable peak height in the evolved wave function (see Fig.~\ref{fig:messwave}), we may obtain a huge variance in the phase shifts or even wrong values. 
	In many cases, we notice the phase shifts computed by the difference between the free plane wave and the evolved wave function are correct, as shown in Table~\ref{tab:phaseshift_probability}. 
	
	Applying the variational method of Sec.~\ref{sec:variational} allows us to obtain the correct phase shift even with noisy wave function. 
	To prove it, we approximate that each front of the evolved wave function has a different amplitude $\{A_i\}$, but with the same phase $\delta_L$ and momentum $k_{\rm in}$. As discussed in the main text, this was the case of Fig.~\ref{fig:messwave}. Under this approximation, we write the evolved wave function as
	
	\begin{equation}
		\psi(t,r)=\sum_i A_i \,{\rm Box}\left(r^i_1<r<r^i_2\right)\sin(k_{\rm in}r+\frac{L\,\pi}{2}+\delta_L)\,
	\end{equation}
	
	where $\rm{Box}$ indicates the box function from $r^i_1=\frac{2\pi n_i+\delta_L}{k}$ to $r^i_2=\frac{2\pi (n_i+1) +\delta_L}{k}$.
	
	Using Eq.~\eqref{eq:scatteringeq}, we obtain
	\begin{widetext}
		\begin{equation}
			\begin{split}
				P(\delta_V)&=\left|\int dr\, \sin(k_{\rm in}r+\delta_V+\frac{L\pi}{2}) \psi(t,r)\right|^2 \\
				&=\left|\int dr\,\sin(k_{\rm in}r+\delta_V+\frac{L\pi}{2}) \sum_i A_i\, {\rm Box} \left( r^i_1<r<r^i_2\right) \, \sin(k_{\rm in}r+\delta_L+\frac{L\pi}{2})\right|^2\\
				& =\sum_i A_i^2 \cos^2(\delta_V-\delta_L)\,.\\
			\end{split}\label{eq:proof_variational}
		\end{equation}
	\end{widetext}

	We can see that, applying the TEPS algorithm (see Sec.~\ref{sec:realtime}), we may have some errors due to the different $A_i$ values. Indeed, the TEPS phase shift is given by:
	\begin{equation}
		\delta^0_L=\arccos\left(c_0\, \sqrt{\sum_i A_i(t)^2 \cos^2(\delta_L)} \right)\,.
	\end{equation}
	Usually, when all the amplitudes are constant, $c_0$, the ratio of normalizations, cancels the $A$ contribution. Otherwise, with different amplitudes $\{A_i(t)\}$, we can obtain an uncorrected overlap probability, $\frac{\sum_i |A_i(t)|^2 }{N\,A^2} \cos^2(\delta_L)$. Hence, our phase shift result has a big variance in time, or in the worst case, it may be even  wrong.
	
	In the variational method, the maximum position is $\delta_V=\delta_L$ because it is independent of the amplitudes but only from the cosine argument.

	\section{IBM quantum circuits \label{app:ibm_quantum_circuit}}
	
	As discussed in the main text and as shown in Fig.~\ref{fig:qc_truncation}, each quantum circuit of the V-TEPS algorithm is divided into three main gates. The first gate, $G(k_{\rm in})$ prepares the qubit wave function into a truncated plane wave. This gate is kept the same in all the quantum circuits. The second gate evolves at time $t$ the state and $D^\dagger$ gate applies the detector state (see Eq.~\eqref{eq:detector_state_box}). In the first step of V-TEPS, we change the time in the real-time operator, keeping the detector gate in the same setting $\delta_V=0$. In the second time step, to find the correct time $t$, we fix the time on the real-time operator and compute the detector gate for each $\delta_V$. 
	
\begin{figure*}[ht]
		\subfloat[Initial state preparation quantum circuit.\label{fig:reinitialization_qc}] {\includegraphics[width=\textwidth]{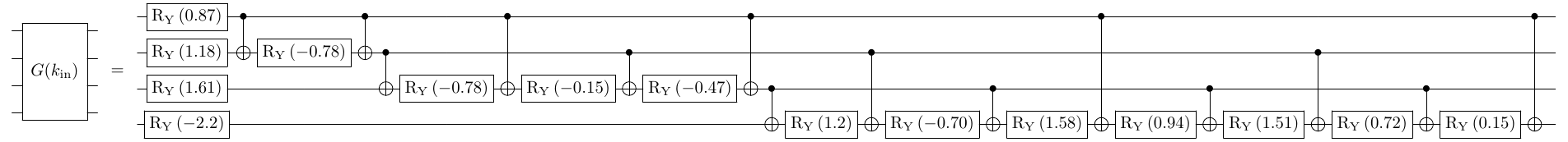}}\\
		\subfloat[Real time evolution quantum circuit for $t=600$ MeV$^{-1}$. \label{fig:realtime_qc}]{\includegraphics[width=\textwidth]{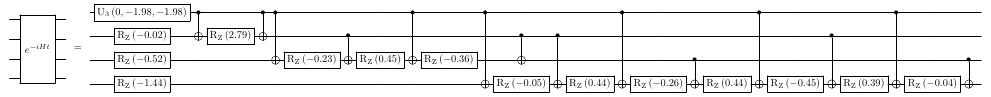} }\\
		\subfloat[Detector quantum circuit setting $\delta_V=0$. \label{fig:detector_qc}]{\includegraphics[width=\textwidth]{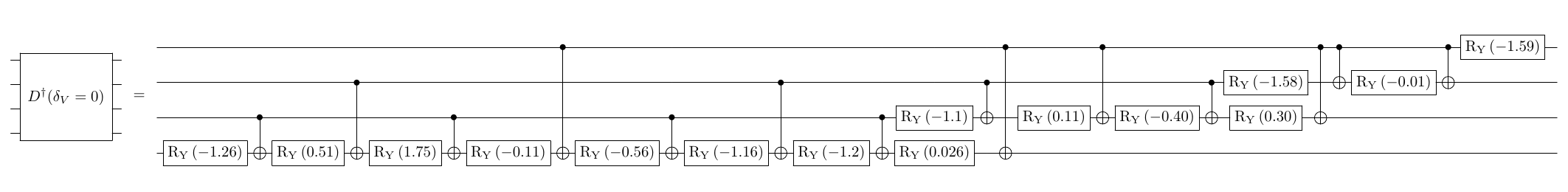} }\\
		\subfloat[Detector quantum circuit setting $\delta_V=-\frac{\pi}{2}$. \label{fig:detector_qc2}]{\includegraphics[width=\textwidth]{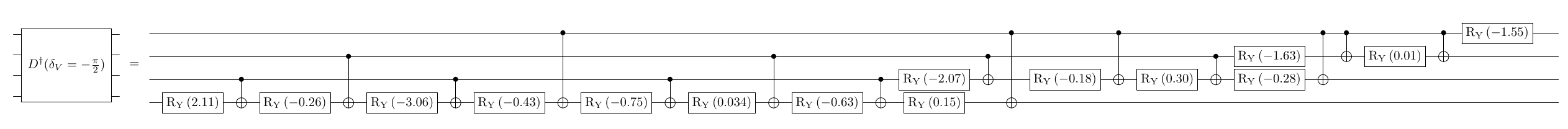} }\\
		\label{fig:IBM_decomposition}
		\caption{Implemented quantum circuits}
	\end{figure*}

	In our simulation, we apply the state-preparation algorithm of Ref.~\cite{PhysRevA.102.012612} to compile the $G(k_{\in})$ and $D^\dagger$ gate. Fig.~\ref{fig:reinitialization_qc} shows our compilation of $G(k_{\rm in})$ in a digital gate set $\{R_y, R_z, U_3, CNOT\}$. Fig.~\ref{fig:detector_qc} and \ref{fig:detector_qc2} show the detector decomposition when we set $\delta_V=0$ and $\delta_V=-\frac{\pi}{2}$, respectively. we can observe that the gate structure derives from going from right to left in Fig.~\ref{fig:reinitialization_qc}. We compute the $R_y$ angle sequence for detector states with a different $\delta_V$. Fig.~\ref{fig:realtime_qc} shows the real-time operator for $t=600$ MeV$^{-1}$ (the time used in the second step of V-TEPS in Sec~\ref{sec:results_quantum_hardware}). The $R_z$ angles are proportional to the implemented time $t$.
	
	Practically, to evaluate the angle in the state preparation, we classically compute the $R_y$ angles that give us the following states
	\begin{equation}
		\begin{pmatrix}
			\braket{n_0}{\phi_T}\\
			\braket{n_1}{\phi_T}\\
			...\\
			\braket{n_{15}}{\phi_T}\\
		\end{pmatrix}
	\end{equation}
	where $\ket{n_i}$ indicates the $i$-th eigenstate of the Hamiltonian and $\ket{\phi_T}$ the target state, that is, the initial truncated plane wave or the detector state with the phase $\delta_V$,
	\begin{equation}
		\phi_T = \frac{1}{1+e^{-\frac{(r-110)}{20}}} \,j_L(k_{\rm in} r ) 
	\end{equation}
	or
	\begin{equation}
		\phi_T = {\rm Box}(r) \,j_L(k_{\rm in} r ) \,.
	\end{equation}
	
	\section{Decoherence Renormalization method \label{app:DR}}
	
	Applying the Decoherence Renormalization method allows us to rescale the probability results considering the effect of decoherence. For each desired quantum circuit, we run two different quantum circuits: the first one implements the original quantum circuit, and the second one runs a quantum circuit with the same structure as the original circuit, except its action on the initial state is the identity. In this work, the identity gate is obtained by imposing all the single qubit angles equal to 0. This identity quantum circuit allows us to estimate the scaling factor due to decoherence.
	Indeed, we can estimate the theoretical probability of the original quantum circuit ($P_{phys}^{\rm ex}(t)$) using the following expression~\cite{Decoherence_renorm}
	\begin{equation}
		P_{phys}^{\rm ex}(t)-1/2^n = \frac{ P_{id}^{\rm ex}-1/2^n}{    P_{id}^{\rm noisy}-1/2^n
		} \left( P_{phys}^{\rm noisy}(t)-1/2^n\right) \,,
		\label{eq:DR}
	\end{equation}
	where $n$ indicates the number of qubits, $P_{phys}^{\rm noisy}(t)$ indicates the obtained probability value from the original quantum circuit, and $P_{id}^{\rm noisy}$ indicates the obtained probability from the identity quantum circuit. $P_{id}^{\rm ex}$ corresponds to the noiseless result of the identity quantum circuit, in our case, $P_{id}^{\rm ex}=1$.
\end{document}